\documentclass[twocolumn]{aastex63}
\usepackage{amsmath}
\usepackage{xcolor}
\received{April 10, 2023}
\revised{July 25, 2023}
\accepted{July 26, 2023}
\submitjournal{ApJ}

\shorttitle{Universal uIMF of the LEGUS star clusters}
\shortauthors{Jung et al.}

\graphicspath{{./}{figures/}}

\begin{document}

\title{Universal Upper End of the Stellar Initial Mass Function in the Young and Compact LEGUS clusters}

\correspondingauthor{Dooseok Escher Jung}
\email{djung@umass.edu}

\author[0000-0003-2797-9979]{Dooseok Escher Jung}
\affiliation{
Department of Astronomy, 
University of Massachusetts Amherst, 
710 North Pleasant Street, 
Amherst, MA 01003, USA
}

\author[0000-0002-5189-8004]{Daniela Calzetti}
\affiliation{
Department of Astronomy, 
University of Massachusetts Amherst, 
710 North Pleasant Street, 
Amherst, MA 01003, USA
}

\author[0000-0003-1427-2456]{Matteo Messa}
\affiliation{
Observatoire de Genève,
Université de Genève,
51 Ch. des Maillettes,
CH-1290 Versoix, Switzerland
}

\author[0000-0002-3871-010X]{Mark Heyer}
\affiliation{
Department of Astronomy, 
University of Massachusetts Amherst, 
710 North Pleasant Street, 
Amherst, MA 01003, USA
}

\author[0000-0002-0923-8352]{Mattia Sirressi}
\affiliation{
Department of Astronomy, Oscar Klein Centre,
Stockholm University, AlbaNova, SE-106 91 Stockholm, Sweden
}

\author[0000-0002-1000-6081]{Sean T. Linden}
\affiliation{
Department of Astronomy, 
University of Massachusetts Amherst, 
710 North Pleasant Street, 
Amherst, MA 01003, USA
}

\author[0000-0002-8192-8091]{Angela Adamo}
\affiliation{
Department of Astronomy, Oscar Klein Centre,
Stockholm University, AlbaNova, SE-106 91 Stockholm, Sweden
}

\author[0000-0003-0085-4623]{Rupali Chandar}
\affiliation{
Ritter Astrophysical Research Center,
University of Toledo, Toledo, OH 43606, USA
}

\author[0000-0001-6291-6813]{Michele Cignoni}
\affiliation{
Department of Physics, University of Pisa,
Largo Pontecorvo, 3, I-56127 Pisa, Italy
}

\author[0000-0002-6877-7655]{David O. Cook}
\affiliation{
Caltech/IPAC, 1200 E. California Boulevard,
Pasadena, CA 91125, USA
}

\author[0000-0002-4578-297X]{Clare L. Dobbs}
\affiliation{
School of Physics and Astronomy,
University of Exeter, Stocker Road,
Exeter EX4 4QL, UK
}

\author[0000-0002-1723-6330]{Bruce G. Elmegreen}
\affiliation{
IBM T. J. Watson Research Center,
1101 Kitchawan Road, Yorktown Heights,
NY 10598, USA
}

\author[0000-0003-2638-1334]{Aaron S. Evans}
\affiliation{
Department of Astronomy,
University of Virginia, 530 McCormick Road, Charlottesville, VA 22904, USA
}

\author[0000-0001-6676-3842]{Michele Fumagalli}
\affiliation{
Dipartimento di Fisica G. Occhialini,
Universit\`a degli Studi di Milano-Bicocca, Piazza della Scienza 3,
20126 Milano, Italy 
}
\affiliation{
INAF - Osservatorio Astronomico di Trieste, via G. B. Tiepolo 11,
34143 Trieste, Italy
}

\author[0000-0001-8608-0408]{John S. Gallagher III}
\affiliation{
Department of Astronomy,
University of Wisconsin,
475 N. Charter Street, Madison,
WI, 53706, USA
}

\author[0000-0002-3322-9798]{Deidre A. Hunter}
\affiliation{
Lowell Observatory,
1400 West Mars Hill Road, Flagstaff,
AZ 86001, USA 
}

\author[0000-0001-8348-2671]{Kelsey E. Johnson}
\affiliation{
Department of Astronomy,
University of Virginia, 530 McCormick Road,
Charlottesville, VA 22904, USA
}

\author[0000-0001-5448-1821]{Robert C. Kennicutt, Jr.}
\affiliation{
Department of Astronomy and Steward Observatory, University of Arizona, Tucson, AZ 85721, USA
}
\affiliation{
Department of Physics and Astronomy,
Texas A\&M University, College Station,
TX 77843, USA
}
\affiliation{
Institute of Astronomy,
University of Cambridge, Madingley Road,
Cambridge CB3 0HA, UK
}

\author[0000-0003-3893-854X]{Mark R. Krumholz}
\affiliation{
Research School of Astronomy and Astrophysics,
Australian National University, Canberra,
ACT 2601, Australia 
}

\author[0000-0001-7144-7182]{Daniel Schaerer}
\affiliation{
Observatoire de Genève,
Université de Genève, Chemin Pegasi 51,
1290 Versoix, Switzerland
}
\affiliation{
CNRS, IRAP, 14 Avenue E. Belin,
31400 Toulouse, France
}

\author[0000-0003-2954-7643]{Elena Sabbi}
\affiliation{
Space Telescope Science Institute,
3700 San Martin Drive, Baltimore,
MD, 21218, USA
}

\author[0000-0002-0806-168X]{Linda J. Smith}
\affiliation{
Space Telescope Science Institute,
3700 San Martin Drive, Baltimore,
MD, 21218, USA
}

\author[0000-0002-0986-4759]{Monica Tosi}
\affiliation{
INAF – OAS Osservatorio di Astrofisica e Scienza dello Spazio,
I-40129 Bologna, Italy
}

\author[0000-0001-8289-3428]{Aida Wofford}
\affiliation{
Instituto de Astronomia, Universidad Nacional Autonoma de Mexico,
Unidad Académica en Ensenada,
22860 Ensenada, Mexico
}

\begin{abstract}

We investigate the variation in the upper end of stellar initial mass function (uIMF) in 375 young and compact star clusters in five nearby galaxies within $\sim 5$~Mpc. All the young stellar clusters (YSCs) in the sample have ages $\lesssim 4$~Myr and masses above 500 $M_{\odot}$, according to standard stellar models. The YSC catalogs were produced from Hubble Space Telescope images obtained as part of the Legacy ExtraGalactic UV Survey (LEGUS) Hubble treasury program. They are used here to test whether the uIMF is universal or changes as a function of the cluster's stellar mass. We perform this test by measuring the H$\alpha$ luminosity of the star clusters as a proxy for their ionizing photon rate, and charting its trend as a function of cluster mass. Large cluster numbers allow us to mitigate the stochastic sampling of the uIMF. The advantage of our approach relative to previous similar attempts is the use of cluster catalogs that have been selected independently of the presence of H$\alpha$ emission, thus removing a potential sample bias. We find that the uIMF, as traced by the H$\alpha$ emission, shows no dependence on cluster mass, suggesting that the maximum stellar mass that can be produced in star clusters is universal, in agreement with previous findings.

\end{abstract}


\section{Introduction} \label{sec:intro}

The stellar Initial Mass Function (IMF) is a valuable tool that enables the derivation of physical quantities such as star formation rates (SFRs) \citep{Kennicutt1998} and star formation histories \citep{Tolstoy2009} from calibrations of light emission. Similarly, the assumed IMF affects theoretical predictions of chemical evolution models on galactic scales \citep{Romano2005} and the stellar feedback into the interstellar/circumgalactic media \citep{Krumholz2019}. 
To effectively use this tool, it is imperative to establish whether the functional form of the stellar IMF is universal or varies with local or global environments.

\citet{Salpeter1955} was the first to introduce the concept of the stellar IMF, the mass distribution of stars at birth, as a single power-law with a slope $\Gamma \sim 1.35$ in the mass range $\sim 3 \mbox{--} 15 \: M_{\odot}$, in the form:
\begin{equation}
\Phi(\mathrm{log} \, m) = \frac{dN}{d \, \mathrm{log} \, m} \propto m^{-\Gamma}
\end{equation}
where $\mathrm{log} \, m$ is the logarithmic mass of a star and $N$ is the number of stars between $\mathrm{log} \, m$ and $\mathrm{log} \, m + d \: \mathrm{log} \, m$. 

\citet{Miller1979} found that the slope is shallower at masses lower than those examined by \citet{Salpeter1955}, and fitted the IMF to both a segmented power law and a log-normal function. \citet{Kroupa2001} revised the multi-segmented power law to have $\Gamma = -0.7$ for substellar objects ($M < 0.08 \: M_{\odot}$), $\Gamma = 0.3$ in the lower-stellar mass regime ($0.08 \: \mbox{--} \: 0.5 \: M_{\odot}$) and $\Gamma = 1.3$ for stars with $M > 0.5 \: M_{\odot}$. \citet{Chabrier2003} also fitted the IMF to a log-normal function:
\begin{equation}
\phi(m) \sim \mathrm{exp} (- \, \frac{\mathrm{log} \, m - \mathrm{log} \, m_{c} }{2\sigma^{2}} )
\end{equation}
where $m_{c}$ is a characteristic mass and $\sigma$ is a dispersion of the log-normal distribution.  The Chabrier-type IMF is almost indistinguishable from a Kroupa-type IMF \citep{Dabringhausen2008}. In the context of this paper, therefore, the IMF is termed `universal' if its shape, slopes in the case of the functional form in \citet{Kroupa2001}, and upper/lower stellar limits are independent of the local or global environment. 

The debate on whether or not the IMF depends on other parameters, such as metallicity, environmental properties, etc., has been ongoing for several decades. Some authors argue that the stellar IMF is universal \citep{Massey2003, Bastian2010, Oey2011, Weisz2012}, while  other studies find non-universal IMFs, with variations that are dependent on the galactic environment \citep{Fardal2007, Wilkins2008, VanDokkum2011, Kroupa2012, Cappellari2012, Geha2013}. For example, using star counts, \citet{Schneider2018} found that the slope of the high-mass IMF in 30 Doradus, a giant H II region in the Large Magellanic Cloud (LMC), is shallower than the Salpeter/Kroupa value. Counting stars is generally considered the most accurate method to derive the IMF. 

However, counting stars individually is challenging, especially when studying potential IMF variations in the massive star regime due to the evolution of stars and star clusters. The most massive stars are short lived ($\lesssim 3$~Myr) and their numbers are significantly lower than those of low-mass stars. Moreover, they are mainly found in the dense environments of star clusters, where crowding becomes an issue \citep{Kalari2022}. Mass segregation that occurs in gravitationally bound systems makes more massive stars sink toward the center of the stellar clusters over time, and causes low-mass stars to be dynamically ejected. As a result, in the crowded areas of the cluster centers, massive stars are difficult to distinguish individually, and low-mass stars can remain undetected due to their faintness or the ejection from the cluster itself \citep{MaizApellniz2008, Ascenso2009}. Thus, the problem of crowding complicates attempts at resolving high-mass stars even with the angular resolution of the Hubble Space Telescope (HST). The proximity of the Magellanic Clouds (distances $\sim 50 \mbox{--} 60$~kpc) has helped overcome some of these complications and stars heavier than $\sim 150 \: M_{\odot}$ have been found in the core of R136, the central cluster in 30 Doradus \citep{Crowther2010, Brands2022, Kalari2022}. By using semi-resolved star clusters of intermediate masses ($10^{3} \: M_{\odot} \mbox{--} \: 10^{4} M_{\odot}$) in M31 ($\sim 1$~Mpc), \citet{Weisz2015} found that the slope of the high-mass IMF ($\gtrsim 2 \mbox{--} 3 \: M_{\odot}$) in the clusters is close to the Kroupa value ($\Gamma = 1.45^{+0.03}_{-0.06}$) for the same stellar mass range.

An alternative method to individual star counting consists of measuring the integrated or summed IMF of a young stellar population. This method avoids the issues of the resolved-star approach; however, it comes with challenges of its own. Ideally, the population needs to be very young ($\lesssim 3$~Myr, to contain massive stars), coeval (to provide one representation of the IMF), and homogeneous (to sample the same IMF across the population). Young star clusters (YSCs) provide, at least in principle, such a population. The challenges consist in (1) deriving accurate masses and ages for the star clusters, and (2) deriving the IMF in an unambiguous way from the integrated light of the young cluster population. The latter is challenging even for massive stars, which tend to dominate the light output of YSCs. Furthermore, an accurate census of massive stars can be hindered by feedback, which causes high infant mortality rates in stellar clusters; less than 20\% of cluster systems survive longer than 10 Myr \citep{Lada2003}. A dispersed cluster becomes difficult to recognize as a single-age entity for the purpose of measuring the IMF.

Dispersed clusters contribute their stars to the field and produce an integrated mass function, sometimes called the present-day mass function (PDMF) \citep{Miller1979}. Because high-mass stars disappear before the low-mass stars have scattered away from their birthplace in the galaxy, the PDMF in any limited region has a complicated relationship to the IMF, and they are not equivalent. Also, if the SFR increases with time, then the summed mass function for all the remaining stars will be shallower than the original cluster IMFs \citep{Elmegreen2006}. Even for a steady SFR, and accounting all the stars that ever formed in, e.g., a young, pre-supernova region, the summed IMF from many clusters is not necessarily the same as the IMFs of the individual clusters. They are the same only if stellar mass is randomly sampled from a universal IMF with a power law shape to infinitely high mass, or if the cluster mass function (CMF) has a slope equal to $-2$ or shallower \citep{Elmegreen2006a}.

If the maximum mass of the star(s) depends physically on the cluster mass, increasing with cluster or molecular cloud mass for example \citep{Larson1982, Kroupa2003} and not just randomly by the size of sample effect \citep{Elmegreen1983}, then the summed IMF can be steeper than the individual IMF. This is because a lot of clusters sum  together to yield low-mass stars but only the highest-mass clusters sum  together to yield the high-mass stars. To get purely stochastic sampling (the randomly sampled population of massive stars that can occur in clusters with mass $< 10^{5} \: M_{\odot}$, i.e. with too low a mass to fully sample the IMF) with a summed IMF slope equal to the individual cluster's slope, some low-mass clusters have to produce high-mass stars. Because cluster-forming cloud complexes are usually much more massive than individual stars, there is typically enough gas to do this. Available observations of the Milky Way bulge, large star fields and whole galaxies confirm that the summed IMF from composite star forming regions is about equal to the IMF from individual star forming regions \citep{Elmegreen2006a}.

\citet{Pflamm-Altenburg2009} considered the implications of a physical, rather than stochastic, relation between the mass of the the most massive star ($m_{\mathrm{max}}$) and the mass of the embedded cluster ($M_{\mathrm{ecl}}$) that hosts the star(s) with $m_{\mathrm{max}}$, generally expressed as a $m_{\mathrm{max}} \mbox{--} M_{\mathrm{ecl}}$ relation. Embedded clusters are those clusters that are still retaining their original gas, and are therefore about twice as massive as their exposed counterparts, which we term YSCs \citep{KroupaBoily2002}. \citet{Pflamm-Altenburg2009} developed models for an integrated galactic IMF (IGIMF) to test the $m_{\mathrm{max}} \mbox{--} M_{\mathrm{ecl}}$ relation on galactic scales. They found that this relation can explain observations such as the systematic dearth of H$\alpha$ emission in low-mass galaxies, if low-mass, low-density galaxies mostly form low-mass clusters which, in turn, form a smaller numbers of massive stars than expected from a fully sampled IMF. \citet{Weidner2010} summarized several theoretical studies that predict the $m_{\mathrm{max}} \mbox{--} M_{\mathrm{ecl}}$ relation. In addition, they reported several observational results that show that $m_{\mathrm{max}}$ and $M_{\mathrm{ecl}}$ are correlated in resolved stars and clusters, suggesting that the $m_{\mathrm{max}} \mbox{--} M_{\mathrm{ecl}}$ relation has general validity. \citet{Oey2011} further extended this concept to explain the steepness of the Salpeter IMF in star clusters as the result of an initial clump mass function with a slope of $-2$ combined with an inability of low-mass clumps to form massive stars. Similar considerations can be used to explain the steeper H~II region luminosity function in the interarm regions relative to the spiral arms of galaxies \citep{Oey1998}. \citet{Yan2017} found that the IGIMF model  explains well the correlation between SFR and the mass of the most massive star cluster ($M_{\mathrm {ecl, max}}$) in the  host galaxy, and produce a prediction for the expected supernova rate in very-low SFR galaxies. \citet{Jerabkova2018} applied the IGIMF framework to calculate grids of models that depend on both metallicity and SFR. Their IGIMF model can potentially explain several observational results, including the evolution of the time-varying IMF in massive elliptical galaxies.

Conversely, \citet{Fumagalli2011}, suggested that the IGIMF cannot fully describe the observed large scatter in the H$\alpha$/FUV ratio for low-luminosity galaxies regardless of their SFR. After adding stochasticity and time evolution to a \citet{Kroupa2001} IMF and convolving the IMF with the CMF, \citet{Fumagalli2011} reproduced the observed $L_{\mathrm {H} \alpha} / L_{\mathrm {FUV}}$ ratios in dwarf galaxies without the need to implement the IGIMF framework. Similar results were obtained by \citet{Hermanowicz2013} for the $L_{\mathrm {H} \alpha} / L_{\mathrm {FUV}}$ ratios of a large sample of H~II regions extracted from eight nearby star-forming galaxies. 

In an attempt to discriminate among these different results, \citet{Calzetti2010} used the integrated light from the YSC  population to test the universality of the upper end of the stellar IMF (uIMF), bypassing the traditional method of individual star counts. Based on \citet{Corbelli2009}, individual YSCs were divided in bins of stellar mass and the properties of the clusters were combined together within each mass bin to emulate a high-mass cluster. One of the combined properties is the luminosity of the hydrogen recombination-line intensity, H$\alpha$, as a tracer of the ionizing photon rate, which is a proxy for the number of massive and ionizing stars present in the star clusters located within H II regions \citep{Oey1998}. This method allows one to investigate the universality of the uIMF in star clusters across a wide range of masses, including those below $\sim 3000 \: M_{\odot}$, which are subject to significant stochastic sampling \citep{Cervino2002}. The application of this method to the cluster populations in the nearby galaxies NGC 5194,  NGC 4214 and M~83 \citep{Calzetti2010, Andrews2013, Andrews2014} shows that the clusters with even the lowest mass down to $500 \: M_{\odot}$ produce high-mass stars up to $120 \: M_{\odot}$. Thus, according to these authors, the maximal star mass is not dependent on the mass of its parent cluster, implying a disagreement with a physically-based $m_{\mathrm{max}} \mbox{--} M_{\mathrm{ecl}}$ relation \citep{Weidner2010}.

Intriguingly, \citet{Weidner2014} argue that, in the case of NGC 4214, the addition of stochastic sampling helps reconcile the trends observed in the cluster population of this galaxy with the $m_{\mathrm{max}} \mbox{--} M_{\mathrm{ecl}}$ relation, opposite to the conclusions of \citet{Andrews2013}. However, NGC 4214 alone is not a good test sample for the $m_{\mathrm{max}} \mbox{--} M_{\mathrm{ecl}}$ relation due to its relatively low SFR ($0.2 \: M_{\odot} \, \mathrm{yr}^{-1}$), which implies low numbers of  observed clusters, and thus, low-number statistics. 

Progress on this controversy can be aided by the data products from the Legacy ExtraGalactic UV Survey (LEGUS), a Hubble treasury program that has obtained imaging data  for $\sim 50$ nearby star-forming galaxies (distances $\sim 3.5 \mbox{--} 18$~Mpc) in five bands from NUV to I \citep{Calzetti2015a}. As part of the effort, star cluster catalogs for 31 LEGUS galaxies have been produced, which include visual identification, location and photometry for each star cluster; for clusters detected in more than four bands, physical parameters: age, mass, and color excess, $E(B-V)$, derived from spectral energy distribution fitting \citep{Adamo2017}, are included as well. 

The LEGUS collaboration has already investigated the general properties of the YSC populations in the observed galaxies, including the CMF and Cluster Luminosity Functions (CLF). \citet{Adamo2017}, \citet{Messa2018a} and \citep{Cook2019} generalized the $-2$ slope power law behavior of the CMFs and CLFs to a much larger sample of galaxies than previously analyzed including dwarf galaxies. \citet{Adamo2017} and \citet{Messa2018a, Messa2018b} determined the value of the upper mass truncation to the CMF, and investigated its dependency on galactic environment. \citet{Adamo2017} also showed that the different evolutionary timescales between YSCs and compact associations track the hierarchical process of star formation \citep{Grasha2015, Grasha2017a, Grasha2017}. The investigation of the spatial correlation between the LEGUS YSCs and molecular clouds in M51 and NGC 7793 showed that the younger the clusters, the closer they tend to be located to the molecular clouds, spiral arms and the galactic center \citep{Grasha2018, Grasha2019}. \citet{Orozco-Duarte2022} developed a photometric library of synthetic star clusters with stochastically sampled IMFs; by  comparing their models with the young star clusters in  NGC 7793 from the LEGUS catalogs, those authors determine that physical quantities (ages, masses, extinctions) are highly model-dependent and most clusters have multi-peaked age probability distributions.

The present study builds on the results of earlier studies, with the goal of testing whether or not the uIMF varies as a function of the stellar mass of YSCs, using the large cluster catalogs of the LEGUS collaboration \citep{Calzetti2015a, Adamo2017}. We combine the star clusters from  the catalogs of selected LEGUS galaxies with archival H$\alpha$ imaging data to repeat the analysis of \citet{Andrews2013, Andrews2014}. Our approach differs from those papers in that the clusters are selected independently of whether they have detected line emission or not. In this way, we do not bias our analysis, since we do not  exclude low-mass clusters that may be missing massive, ionizing stars due to stochastic sampling. Section \ref{sec:data} explains the data collection and the image processing.  Section \ref{sec:result} describes how we use our data to get to a description of the uIMF. In Section \ref{sec:conclusion}, we discuss our results and their implications.

\section{Data} \label{sec:data}

\begin{figure*}
\centering
\includegraphics[width=1.0\textwidth]{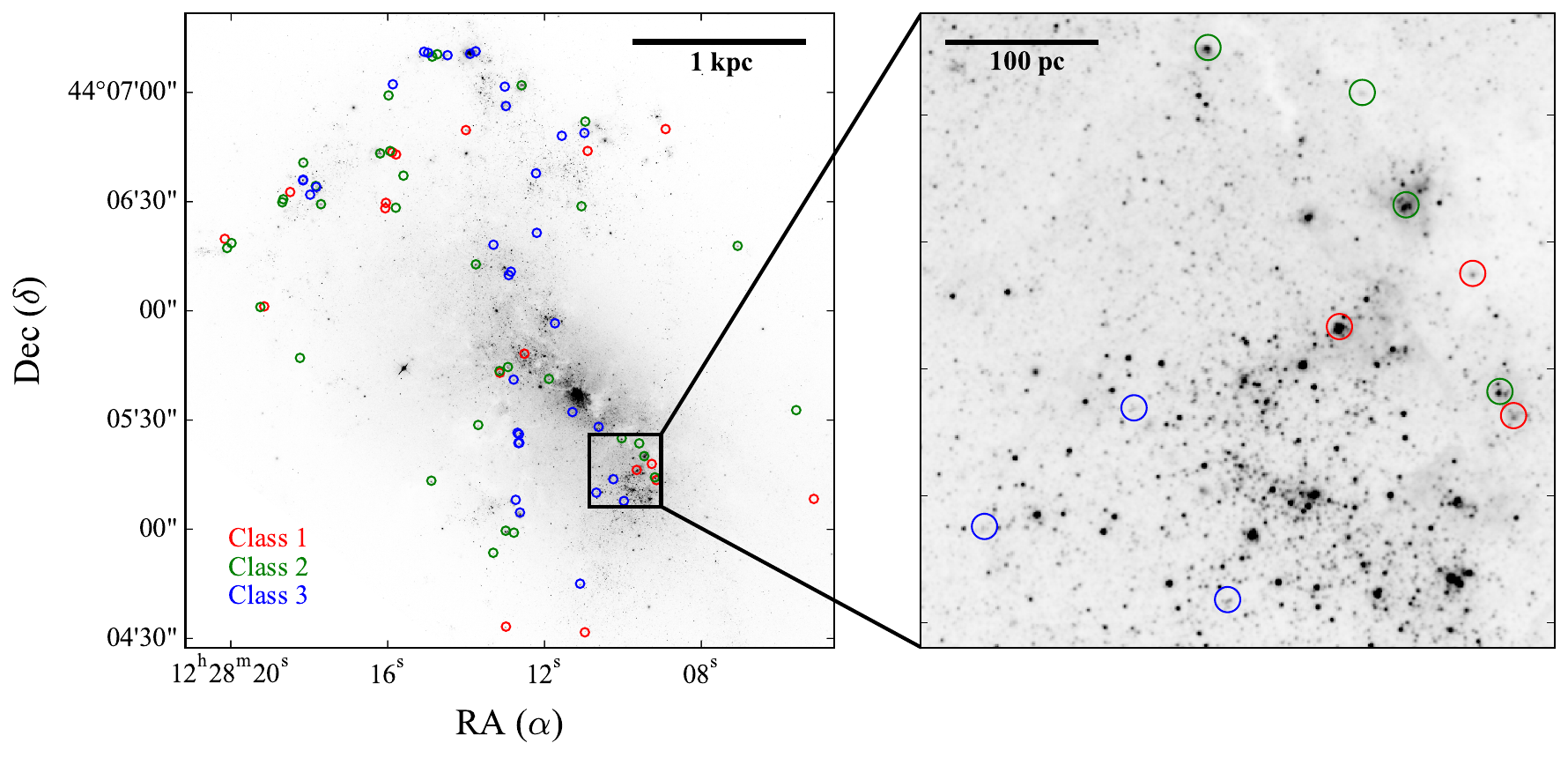}
\caption{\textit{Left}: the locations of the selected LEGUS clusters by class (class 1 as red, class 2 as green and class 3 as blue circles) with ages within 1--4~Myr in NGC 4449 are shown on the F555W image. \textit{Right}: a zoomed-in image of the selected area (the black solid square) in the left panel. Candidates identified as class 4 are excluded here and the morphology-based cluster classification is explained in Section~\ref{sec:selection}.}
\label{fig:class}
\end{figure*}

The Legacy ExtraGalatic UV Survey (LEGUS) is a Hubble Space Telescope (HST) Treasury Program (Cycle 21) which collected five-band images (UV, U, B, V and I) of 50 star-forming galaxies within 12 Mpc with the Wide-Field Camera 3 (WFC3) using the filters F275W, F336W, F438W, F555W and F814W \citep{Calzetti2015a}. Archival images obtained with the Advanced Camera for Surveys (ACS) wide field in the filters F435W, F606W and F814W are also used when they are available to complement the LEGUS observations. The pixel scale of the ACS/WFC is slightly larger, 0$^{\prime\prime}$.05 pixel$^{-1}$, than that of WFC3 (0$^{\prime\prime}$.0396 pixel$^{-1}$), but the LEGUS collaboration aligned and resampled the images to the WFC3 ones.

YSC catalogs were produced using a combination of SExtractor \citep{Bertin1996} for the automatic search of extended sources and human identification. Spectral energy distribution (SED) fits for all sources detected with signal-to-noise (S/N) of 3 or better in at least four bands were performed, and the sources assigned ages, masses, and extinctions, with uncertainties. The YSC catalogs are complete out to $\sim 200$~Myr in age and down to masses between 500 and $5000 \: M_{\odot}$, depending on the age of the source. Details on the catalog production and characteristics can be found in \citet{Adamo2017}. We supplement the LEGUS observations with archival images in the WFC3/F657N or ACS/F658N and the adjacent WFC3/F547M or ACS/F550M (as available) filters to measure the hydrogen recombination line (H$\alpha$).

\subsection{Galaxy and Cluster Selection} \label{sec:selection}
As our goal is to measure the ionizing photon rate of each YSC, we need to separate the ionization regions of neighboring YSCs from each other. This sets an upper limit of about 5 Mpc to the maximum distance of the galaxies whose YSC-produced ionization regions we can resolve with HST. Within this distance limit, individual clusters can be detected down to $\sim 500 \: M_{\odot}$ \citep{Andrews2014} which allows our study to investigate whether massive stars are able to be formed in low-mass clusters. We also require that the YSCs exhibit low dust color excess, $E(B-V)<0.1$ mag to avoid heavy impact from extinction corrections. This choice is likely to select against the youngest (most embedded) YSCs, but it also enables accurate accounting of the ionizing photon rate. The clusters in the LEGUS catalogs show on average low extinction values, typically  $E(B-V) \lesssim 0.2$ mag \citep{Adamo2017, Messa2018a}. After applying these selection criteria, our sample includes five nearby star-forming galaxies with distance $\lesssim$ 5 Mpc. The galaxies span an order of magnitude in  SFRs ($0.1 \mbox{--} 1.2 \: M_{\odot} \: \mathrm{yr}^{-1}$) and  stellar masses ($10^{8} \mbox{--} 10^{9} \: M_{\odot}$); these and other global properties are listed in Table~\ref{tab:object}.

\begin{table*}
\centering
\caption{Properties of the LEGUS galaxies in this study.}
\begin{tabular}{lcccccccc}
\hline\hline
Name$^a$ & RA$^a$ & Dec$^a$ & Morph.$^a$ & Dist.$^b$ (Mpc) & $M_{\bigstar}^{b}$ [$M_{\odot}$] & $E(B-V)^{a}$ & $Z^{c}$ & SFR (UV)$^b$ [$M_{\odot} \: \mathrm{yr}^{-1}$] \\
\hline
NGC 1313 & 03:18:30.05 & -66:28:45.0 & SBd    & 4.39        & 2.6$\times 10^9$ & 0.0965 & 0.008 & 1.15      \\
NGC 1705 & 04:54:13.7  & -53:21:40.9 & SA0/BCG    & 5.1         & 1.3$\times 10^8$ & 0.0071 & 0.008 & 0.11     \\
NGC 4395 & 12:25:56.60 & 33:31:10.1  & SAm    & 4.3         & 6.0$\times 10^8$ & 0.0152 & 0.004 & 0.34     \\
NGC 4449 & 12:28:12.06 & 44:05:42.3  & IBm    & 4.31        & 1.1$\times 10^9$ & 0.0171 & 0.004 & 0.94     \\
NGC 7793 & 23:57:40.86 & -32:35:20.6 & SAd    & 3.44        & 3.2$\times 10^9$ & 0.0171 & 0.008 & 0.52    \\
\hline
\label{tab:object}
\end{tabular}
\footnotesize{$^a$ Galaxy Name, coordinates (J2000), morphological type and the foreground color excess ($A_V / R_V$) due to the dust in our Milky Way, as listed in the NASA Extragalactic Database (NED).

$^b$ Distance, stellar mass and SFR of each galaxy from \citet{Calzetti2015a}.

$^c$ Metallicity of each galaxy, from \citet{Sabbi2018}.}
\end{table*}

The cluster candidates in the LEGUS catalogs have been classified according to four morphological classes as defined in \citet{Adamo2017}. Class 1 candidates are symmetric and concentrated. Class 2 candidates look less symmetric and rather elongated relative to class 1. Class 3 candidates are asymmetric, less compact and have multiple peaks joined by diffuse emission in between the peaks. Candidates in both class 1 and class 2 are considered YSCs while class 3 candidates are ``compact associations". Class 4 candidates are ``non clusters": artifacts, background galaxies, single stars or asterisms. In our analysis, we only use the candidates of classes 1, 2, and 3 as real star clusters; class 4 sources will not be considered any further.

Since we are trying to characterize the uIMF, we require our YSCs to have undergone minimal evolution. We choose clusters with ages $\le$~4 Myr and treat them as coeval objects that have not yet undergone supernova events or  strong gas expulsion \citep{Weidner2010, Relano2012}. Leakage of ionizing photons from H II regions can still occur, and can affect our results, as we discuss below. We require cluster masses above $500 \: M_{\odot}$. This ensures that our YSCs are not dominated by mis-classified single massive stars \citep{Andrews2014}, but are still probing stochastic sampling effects. After all selections are applied, we have 375 clusters in total for this study (Table~\ref{tab:class}). Figure~\ref{fig:class} shows the locations of the clusters identified by class 1, 2, and 3 in NGC 4449.

\begin{table}[b]
\centering
\caption{Statistics of the LEGUS star clusters by class and by galaxy for age within 1--4~Myr, $E(B-V) \leq 0.2$, and mass over $500 \: M_{\odot}$ used in this study.}
\begin{tabular}{lcccc}
\hline\hline
Name & Class 1 & Class 2 & Class 3 & Total \\
 & & & & (each galaxy) \\
\hline
NGC 1313 & 27      & 51      & 88      & 166 \\
NGC 1705 & 0       & 1       & 1       & 2  \\
NGC 4395 & 20      & 42      & 53      & 115 \\
NGC 4449 & 6       & 5       & 2       & 13 \\
NGC 7793 & 15      & 34      & 30      & 79  \\
Total & 68      & 133     & 174    & 375 \\
\hline
\label{tab:class}
\end{tabular}
\end{table}

\subsection{Photometry} \label{sec:phot}

In order to measure the H$\alpha$ luminosity of each star cluster, we retrieve from the Hubble Legacy Archive the images of the five galaxies in the F547M/F550M (V-short) and in the F657N/F658N (H$\alpha$+[N II]) filters obtained with WFC3/ACS. We give preference for the medium band F550M or F547M over the F555W filter to avoid the strong [O III] emission line which affects the broader filter. We align the images to the F814W (I-band) images of the LEGUS data for aperture photometry and additional corrections. Aperture photometry is performed on all 375 clusters in all three bands using the PHOTUTILS package in ASTROPY with a 5 pixel aperture radius, which corresponds to 0.35--0.5 R$_{\rm Str\ddot{o}mgren}$ \citep{Calzetti2010} and using an annulus between 7 to 9 pixel radius for local background. The positions of the selected clusters are taken from the LEGUS catalogs.

The flux density in each band is obtained by converting counts to erg$\cdot$cm$^{-2} \cdot$s$^{-1}$ after local background subtraction. Flux densities in the H$\alpha$ in each cluster are measured from the difference between the observed flux density in the H$\alpha$+[N II] filters (F657N/F658N) and that of the underlying stellar continuum in the narrow band obtained from the interpolation between the F547M/F550M and the F814W filters. The flux conversion keyword PHOTFLAM from the images' header is then used to convert the counts to physical units. Finally, we convert the flux densities to fluxes in H$\alpha$+[N II] by multiplying our measurements by the FWHMs of the narrow-band filters. In the case of undetected H$\alpha$, the detection limits of each cluster in the F657N/F658N filters are reported as upper limits.

Aperture corrections are applied to all photometry measurements. The continuum subtracted H$\alpha$+[N II] (F657M/F658M) images were used to select isolated sources with strong emission to construct curves of growth. The final aperture correction is derived from the average calculated for the isolated sources. The typical correction is a factor $\sim 2.3$. As our sources are all extremely young, the nebular emission, when present, is generally concentrated and coincident with the location of the star cluster's stellar emission. Thus, we do not find large variations in the correction factor. Removal of the contamination by [N II] in the F657M/F658M filters is performed by applying the average ratios of [N II] to  H$\alpha$ of each galaxy to the flux densities in H$\alpha$+[N II] \citep{KennicuttJr.2008}.

\begin{figure*}
\centering
\includegraphics[width=1.0\textwidth]{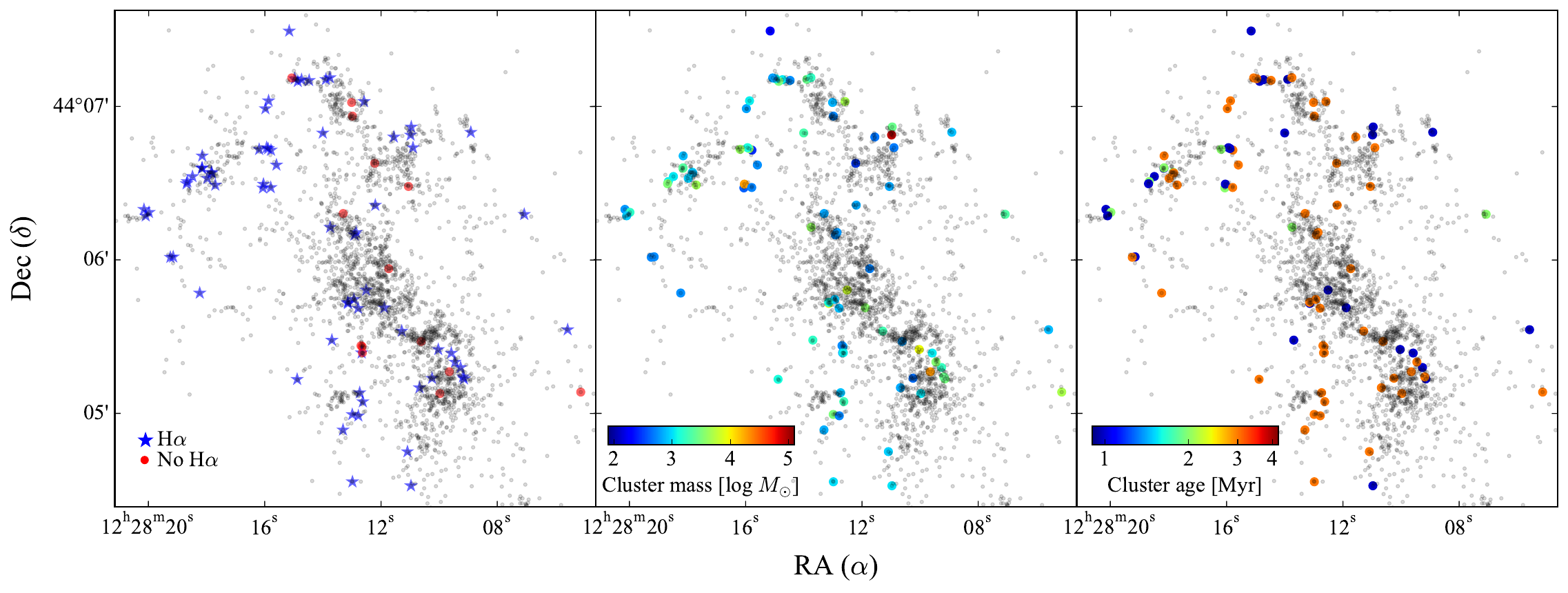}
\caption{The spatial distribution of the star clusters in NGC 4449 selected for this study, and showing the presence/absence of H$\alpha$ (\textit{left}), their masses (\textit{middle}) and their ages (\textit{right}). The values of the masses ($10^{2} \mbox{--} 10^{5} \: M_{\odot}$) and the ages (1--4~Myr) are shown by the color bars in the \textit{middle} and \textit{right} panels, respectively. The gray dots are all star clusters in NGC 4449 from the LEGUS catalogs.}
\label{fig:distribution}
\end{figure*}

Extinction correction for $L_{\mathrm{H} \alpha }$ is performed considering two contributions; the foreground Milky Way color excess, $E(B-V)_{\mathrm {MilkyWay}}$, and the color excess intrinsic to the cluster, $E(B-V)_{\mathrm {cluster}}$. The latter is determined from the SED fitting of the cluster's photometry \citep{Adamo2017}. For the intrinsic attenuation of the nebular gas, we adopt the recipe of \citet{Calzetti1994} and \citet{Calzetti2000}, which establish: $E(B-V)_{\mathrm {gas}} = E(B-V)_{\mathrm {cluster}} / 0.44$. The attenuation affecting the star cluster can then be expressed as:
\begin{equation}
F_{\mathrm {int}} (\lambda) = F_{\mathrm {obs}} (\lambda) \: 10^{\, 0.4 \, E(B-V) \, k(\lambda)}
\end{equation}
where $k(\lambda)=2.54$ for H$\alpha$ and $E(B-V) = E(B-V)_{\mathrm{Milky Way}} + E(B-V)_{\mathrm{cluster}} / 0.44$. Figure~\ref{fig:distribution} shows an example of the spatial distribution of the selected clusters for this study in NGC 4449, indicating whether the H$\alpha$ emission is present or absent, and assigning color-coding to their ages and  masses.

\section{Analysis and results} \label{sec:result}

The number of massive stars in a cluster can be inferred from the hydrogen ionizing photon rate, $Q(\mathrm{H}^{0})$, due to the strong ionizing emission from young and hot OB stars. The H$\alpha$ luminosity, $L_{\mathrm{H} \alpha}$, of a cluster is a tracer of $Q(\mathrm{H}^{0})$, such that we can test whether or not the uIMF depends on cluster mass by normalizing $L_{\mathrm{H} \alpha}$ by the cluster mass ($M_{\mathrm{cl}}$) \citep{Calzetti2010}.

\begin{figure*}
\centering
\includegraphics[width=1.0\textwidth]{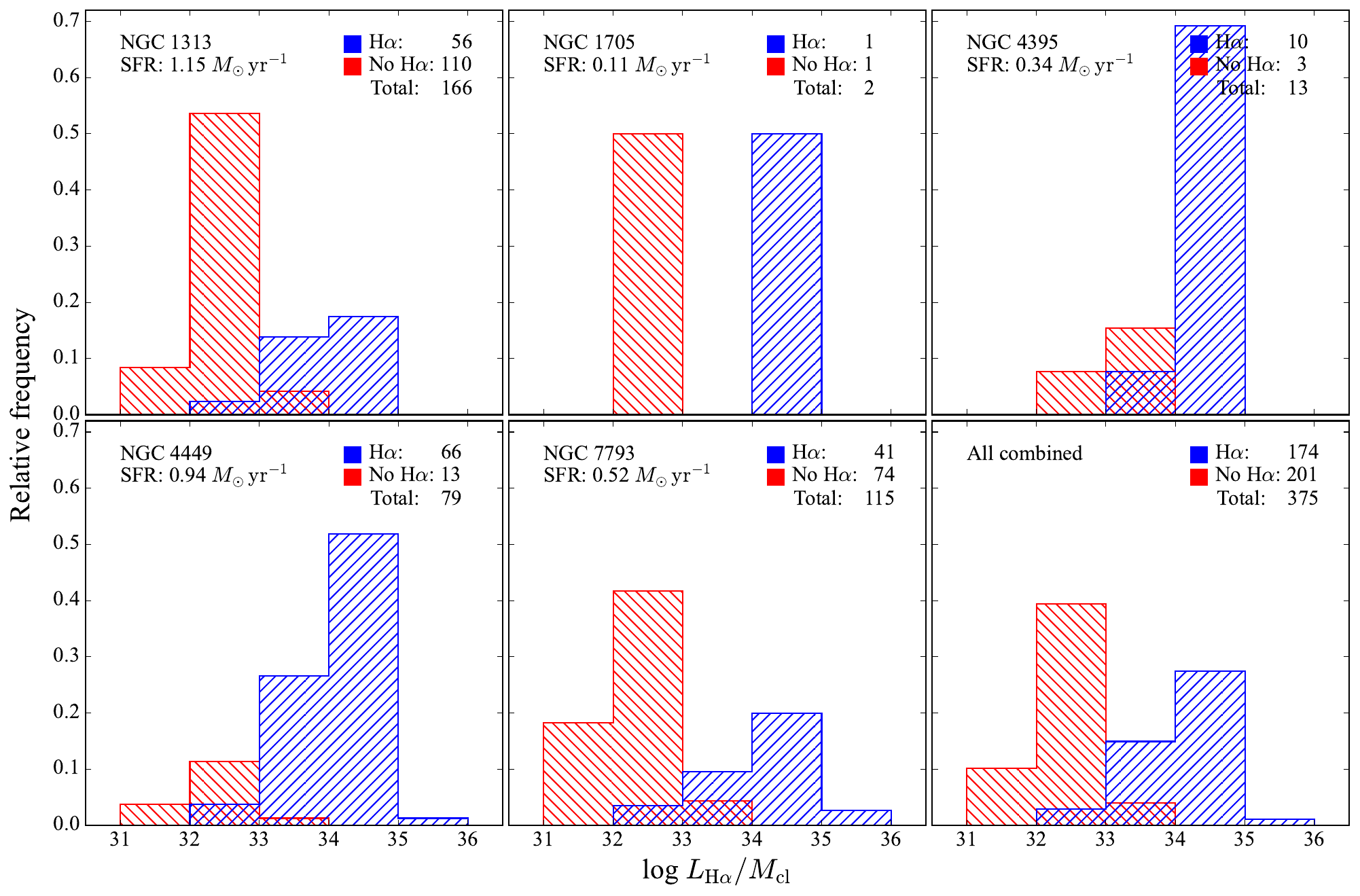}
\caption{The distributions of $L_{\mathrm{H} \alpha} / M_{\mathrm{cl}}$ values for each galaxy and for all samples combined. The blue bars are H$\alpha$-emission detected sources and the red bars are undetected ones, for which the indicated values are upper limits. For the sources detected in H$\alpha$ the typical 2-sided 1$\sigma$ uncertainties are about 1/2 bin in size. The SFR of each galaxy is given  under the name of the galaxy.}
\label{fig:histogram}
\end{figure*}

\begin{figure*}
\centering
\includegraphics[width=1.0\textwidth]{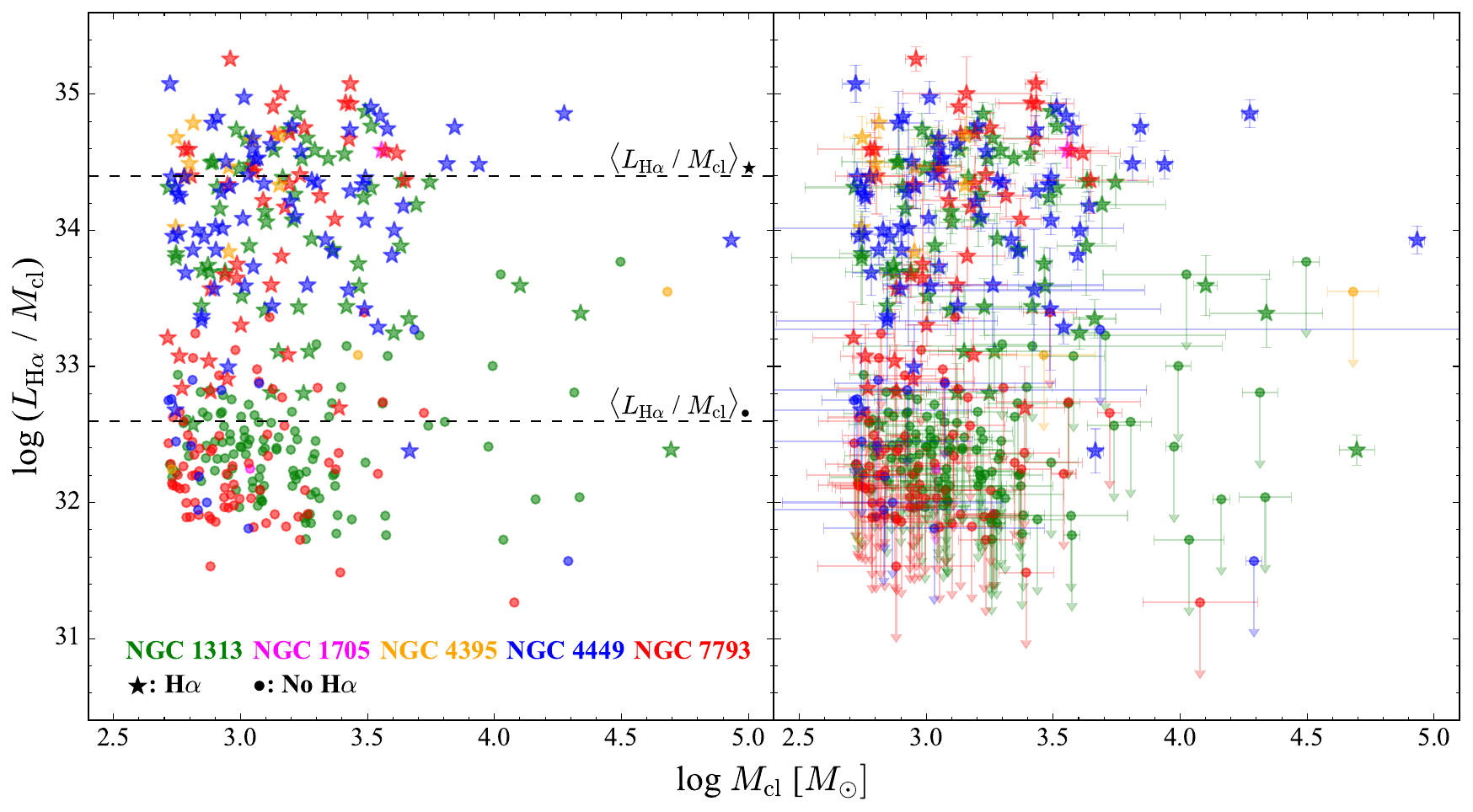}
\caption{\textit{Left}: the ratio $L_{\mathrm{H} \alpha} / M_{\mathrm{cl}}$ of all the selected clusters as a function of stellar  mass, as listed in the LEGUS catalogs. Star markers are the clusters with H$\alpha$ emission, and dots are those without H$\alpha$ emission. For these, the indicated values of $L_{\mathrm{H} \alpha}$ are upper limits. Dashed horizontal lines are the average of $L_{\mathrm{H} \alpha} / M_{\mathrm{cl}}$ for the sources detected in H$\alpha$ ($\langle L_{\mathrm{H} \alpha} / M_{\mathrm{cl}} \rangle _{\bigstar}$) and those undetected in the line ($\langle L_{\mathrm{H} \alpha} / M_{\mathrm{cl}} \rangle _{\bullet}$), respectively. \textit{Right}: the same plot as the left panel, but including the error bars of each cluster. Here the upper limits are clearly marked.}
\label{fig:individual}
\end{figure*}

The distributions of the ratio $L_{\mathrm{H} \alpha} / M_{\mathrm{cl}}$ for our cluster samples are shown as histograms of the relative frequencies for individual galaxy and for all samples combined in Figure~\ref{fig:histogram}; we plot separately the H$\alpha$ detected clusters and the non-detected ones as  upper limits. The $\log ( L_{\mathrm{H} \alpha} / M_{\mathrm{cl}} )$ values cover the range 31--35 when all 375 YSCs are considered. The ratio $L_{\mathrm{H} \alpha} / M_{\mathrm{cl}}$ of the selected clusters in the five galaxies are shown individually in Figure~\ref{fig:individual}. We can hardly see any clear trend between $L_{\mathrm{H} \alpha} / M_{\mathrm{cl}}$ and $M_{\mathrm{cl}}$, suggesting a weak correlation between $m_{\mathrm{max}}$ and $M_{\mathrm{cl}}$.

The clusters with undetected  H$\alpha$ emission represent 54\% of the total number of sources and are located $\sim 1.8$ orders of magnitude on average below the sources with H$\alpha$ emission. Given that the majority of our star clusters have masses below $\sim 3000 \: M_{\odot}$ (Figure~\ref{fig:individual}), the large number of sources without H$\alpha$ emission is easily explainable as an effect of the stochastic sampling of the IMF in this regime. Thus, Figure~\ref{fig:histogram} supports  that the LEGUS sample selection is unbiased relative to presence or absence of nebular emission. The gap between the two distributions is likely an effect of our measurement technique coupled with evolutionary effects of the clusters. For instance, very young clusters hosting massive stars can develop energetic stellar winds over short timescales that remove the nebular gas surrounding the cluster \citep{mackenty2000, Krumholz2019}. Thus, we cannot exclude that, at least for some of the sources without detected H$\alpha$ emission, the lack of detection is due to purely morphological effects because the H$\alpha$ emission may actually be present but located in a shell around the cluster \citep{mackenty2000, Calzetti2010, Relano2012, Andrews2013, Calzetti2015b, Belfiore2022}. An extreme example is the case of  young, UV-bright starbursts with high SFRs and many massive O stars but which lack of ionized gas due to ionizing photon leakage \citep{Marques-Chaves2022}. In this case, our photometric measurements provide a lower limit to the true nebular emission. We cannot correct for this effect as most of our sources are located in crowded regions, where their H$\alpha$ emission overlaps with that of neighboring sources and shells of nebular emission cannot be easily identified and measured.

Conversely, we expect many of the low-mass sources undetected in H$\alpha$ to be subject to stochastic sampling effects, meaning that their low mass prevents them from forming massive stars in significant numbers. In these cases, the sources cannot ionize the surrounding gas and our reported H$\alpha$ values are upper limits. As we cannot discriminate, for the H$\alpha$-undetected sources, between the upper limits from stochastic sampling and the lower limits from expelled gas (see above), we will consider both cases in the analysis that follows. 

\begin{figure*}
\centering
\includegraphics[width=1.0\textwidth]{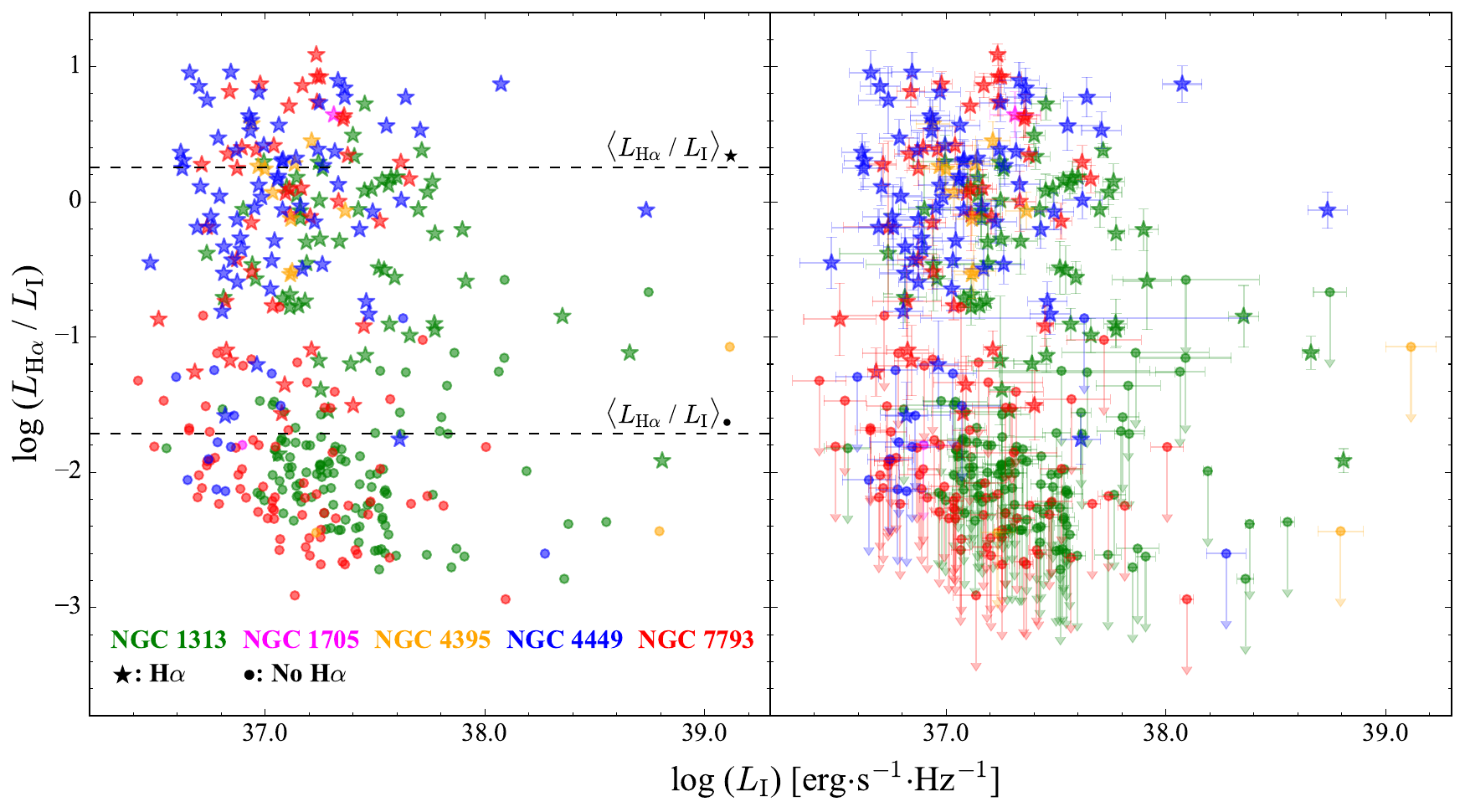}
\caption{The ratio $L_{\mathrm{H} \alpha} / L_{\mathrm{I}}$ of the cluster samples in this study as a function of $L_{\mathrm{I}}$. This figure has the same structure with Figure~\ref{fig:individual} but $L_{\mathrm{I}}$ instead of $M_{\mathrm{cl}}$.}
\label{fig:individual_i}
\end{figure*}

Before binning by mass to mitigate stochasticity, we check whether the physical parameter we use, the cluster mass, may be influenced by the way the SED fitting performed on the photometry of the clusters in the LEGUS catalogs \citep{Adamo2017}. In other words, we want to make sure that the values of the cluster masses are not biased by potential covariances in the SED fitting results due to photometric uncertainties. In order to perform this assessment, we carry out the same procedure as in Figure~\ref{fig:individual} but using  purely observational quantities, i.e., the I-band (F814W) flux density instead of $M_{\mathrm{cl}}$. The choice of this band is driven by its low sensitivity to extinction. Figure~\ref{fig:individual_i} shows the distribution of the clusters' $L_{\mathrm{H} \alpha} / L_{\mathrm{I}}$ as a function of $L_{\mathrm{I}}$. Figure~\ref{fig:individual_i} has, in the low-luminosity regime ($L_{\mathrm{I}} < 38$), a comparable scatter to Figure~\ref{fig:individual} in the low-mass regime; there is a gap between the averages of $L_{\mathrm{H} \alpha} / L_{\mathrm{I}}$ for sources detected and undetected in H$\alpha$, of comparable magnitude to the gap in Figure~\ref{fig:individual}. The errors in  $L_{\mathrm{H} \alpha} / L_{\mathrm{I}}$ do not show a trend as a function of  $L_{\mathrm{I}}$, again similar to the data in Figure~\ref{fig:individual}. This  suggests that the clusters' $L_{\mathrm{H} \alpha}$ and their masses are not strongly correlated with each other, and there is no bias in using the cluster mass as our reference parameter.

We test the $m_{\mathrm{max}} \mbox{--} M_{\mathrm{ecl}}$ relation with the approach suggested by \citet{Calzetti2010}. The IMF and the IGIMF are equivalent under an assumption of a universal IMF. The only effect at play is stochastic sampling, which is predominant at low cluster masses  \citep{Hermanowicz2013}. A low-mass cluster will be missing stars at randomly sampled masses; since, for a typical IMF functional shape, there are far more low-mass stars than high-mass stars, random sampling will affect the latter more than the former. In order to be truly stochastic, the incomplete sampling needs to be erased when accumulating multiple low-mass clusters together. That is, the shape of the IMF resulting from the sum of 100 individual clusters with each mass of $1000 \: M_{\odot}$ will be equivalent to the IMF of a single $10^5 \: M_{\odot}$ cluster. \citep{Elmegreen2000, Elmegreen2006a, Bastian2010, Calzetti2010}. So, we sum all of the values of $L_{\mathrm{H} \alpha} / M_{\mathrm{cl}}$ as
\begin{equation}
\langle \frac{L_{\mathrm{H} \alpha}}{M_{\mathrm{cl}}} \rangle = \frac{ \Sigma_{i} \, L_{\mathrm{H} \alpha, i}}{\Sigma_{i} \, M_{\mathrm{cl}, i}},
\label{eq:ratio_avg}
\end{equation}
which we divide into several mass bins of equal total mass each  (Table~\ref{tab:bin}) to test any trend of the uIMF with mass \citep{Andrews2013, Andrews2014}. We include in the sum above also clusters that are undetected in H$\alpha$ to avoid biasing our results \citet{Weidner2014}. 

\begin{figure*}
\centering
\includegraphics[width=1.0\textwidth]{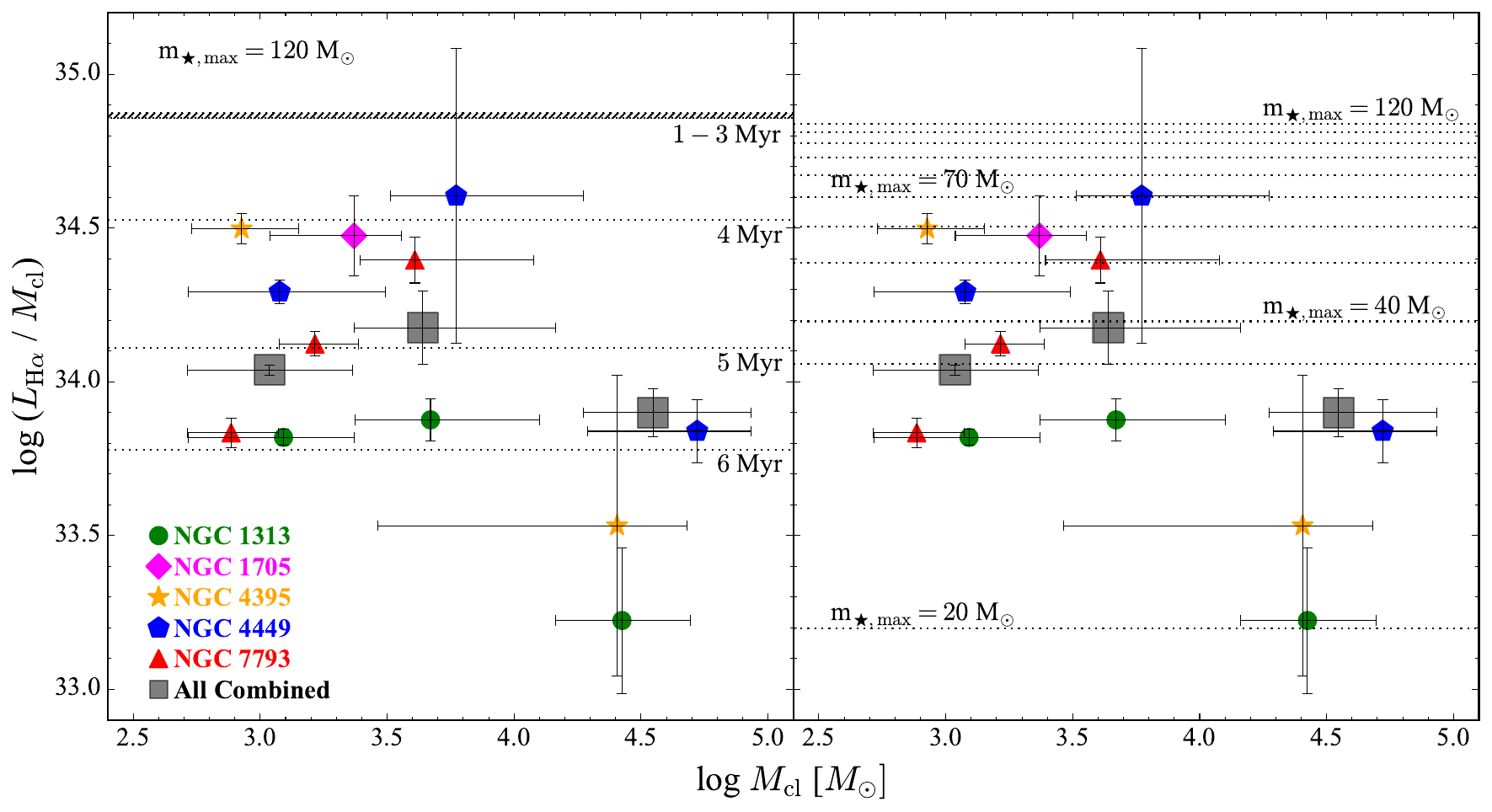}
\caption{\textit{Left}: the location of $\langle L_{\mathrm{H} \alpha} / M_{\mathrm{cl}} \rangle$ ($= \Sigma \, L_{\mathrm{H} \alpha} \, / \, \Sigma \, M_{\mathrm{cl}}$) with their standard deviations (Table~\ref{tab:bin}) as a function of the mean of the samples' mass in each bin. Different colors and shapes identify different galaxies. The dark-gray squares are the $\langle L_{\mathrm{H} \alpha} / M_{\mathrm{cl}} \rangle$ ratios of all galaxies combined, as a function of binned mass. The horizontal bar associated with each symbol shows the size of each mass bin. The horizontal dotted lines show the evolution of model $\langle L_{\mathrm{H} \alpha} / M_{\mathrm{cl}} \rangle$ as a function of age (1--6~Myr) with  fixed $m_{\mathrm{max}} = 120 \: M_{\odot}$ from SB99. \textit{Right}: the same data points as the left panel shown with a range of model upper mass values  $m_{\mathrm{max}}$ averaged within the 1--4 Myr age produced by SB99 from $20 \: M_{\odot}$ to $120 \: M_{\odot}$ in $10 \:  M_{\odot}$ intervals (horizontal dotted lines).}
\label{fig:ratio}
\end{figure*}

Figure~\ref{fig:ratio} shows the $\langle L_{\mathrm{H} \alpha} / M_{\mathrm{cl}} \rangle$ values in bins of cluster mass, for each galaxy and for the entire sample combined. Data are divided into several mass bins to check the $m_{\mathrm{max}} \mbox{--} M_{\mathrm{ecl}}$ relation except for NGC 1705, which has only one data point located at $M_{\mathrm{cl}} \sim 10^{3.4} \: M_{\odot}$ due to the small cluster catalog (only two YSCs, Table~\ref{tab:class}). The value of the maximum mass bin in NGC 7793 is $\sim 10$ times lower than those of other galaxies, so $\langle L_{\mathrm{H} \alpha} / M_{\mathrm{cl}} \rangle$ of NGC 7793 does not cover the higher mass range ($M_{\mathrm{cl}} \gtrsim 10^{3.5} \: M_{\odot}$) of the YSCs covered by the other four galaxies.

\begin{table*}
\centering
\caption{Statistics on the binned data by mass in each individual galaxy and in all galaxies combined.}
\begin{tabular}{lcccccc}
\hline\hline
Name \& Bin No. & $\langle M \rangle$ [$M_{\odot}$] & Mass range [$M_{\odot}$] & No. of clusters & $M_{total}$ [$M_{\odot}$] & $\langle L_{\mathrm{H} \alpha} / M_{\mathrm{cl}} \rangle$ & $\sigma_{\langle L_{\mathrm{H} \alpha} / M_{\mathrm{cl}} \rangle}$$^a$ \\
\hline
\multicolumn{1}{c}{\textbf{NGC 1313}} & & & & & & \\
Bin 1 & $1.235 \times 10^{3}$ & $5.211 \times 10^{2} - 2.348 \times 10^{3}$ & 127 & $1.569 \times 10^{5}$ & 33.819 & 0.0256 \\
Bin 2 & $4.685 \times 10^{3}$ & $2.358 \times 10^{3} - 1.260 \times 10^{4}$ & 33 & $1.546 \times 10^{5}$ & 33.876 & 0.0687 \\
Bin 3 & $2.658 \times 10^{4}$ & $1.452 \times 10^{4} - 4.957 \times 10^{4}$ & 6 & $1.595 \times 10^{5}$ & 33.224 & 0.237 \\
\hline
\multicolumn{1}{c}{\textbf{NGC 1705}} & & & & & & \\
Bin 1 & $2.342 \times 10^{3}$ & $1.091 \times 10^{3} - 3.592 \times 10^{3}$ & 2 & $4.683 \times 10^{3}$ & 33.475 & 0.131 \\
\hline
\multicolumn{1}{c}{\textbf{NGC 4395}} & & & & & & \\
Bin 1 & $8.448 \times 10^{2}$ & $5.366 \times 10^{2} - 1.422 \times 10^{3}$ & 11 & $9.292 \times 10^{3}$ & 34.500 & 0.0486 \\
Bin 2 & $2.544 \times 10^{4}$ & $2.902 \times 10^{3} - 4.797 \times 10^{4}$ & 2 & $5.087 \times 10^{4}$ & 33.532 & 0.488 \\
\hline
\multicolumn{1}{c}{\textbf{NGC 4449}} & & & & & & \\
Bin 1 & $1.193 \times 10^{3}$ & $5.205 \times 10^{2} - 3.102 \times 10^{3}$ & 64 & $7.637 \times 10^{4}$ & 34.292 & 0.0388 \\
Bin 2 & $5.911 \times 10^{3}$ & $3.259 \times 10^{3} - 1.881 \times 10^{4}$ & 13 & $7.685 \times 10^{4}$ & 34.604 & 0.480 \\
Bin 3 & $5.260 \times 10^{4}$ & $1.952 \times 10^{4} - 8.567 \times 10^{4}$ & 2 & $1.052 \times 10^{5}$ & 33.839 & 0.103 \\
\hline
\multicolumn{1}{c}{\textbf{NGC 7793}} & & & & & & \\
Bin 1 & $7.684 \times 10^{2}$ & $5.172 \times 10^{2} - 1.181 \times 10^{3}$ & 69 & $5.302 \times 10^{4}$ & 33.834 & 0.0489 \\
Bin 2 & $1.643 \times 10^{3}$ & $1.193 \times 10^{3} - 2.445 \times 10^{3}$ & 33 & $5.422 \times 10^{4}$ & 34.123 & 0.0397 \\
Bin 3 & $4.067 \times 10^{3}$ & $2.471 \times 10^{3} - 1.197 \times 10^{4}$ & 13 & $5.287 \times 10^{4}$ & 34.396 & 0.0745 \\
\hline
\multicolumn{1}{c}{\textbf{All galaxies}} & & & & & & \\
Bin 1 & $1.087 \times 10^{3}$ & $5.172 \times 10^{2} - 2.317 \times 10^{3}$ & 293 & $3.186 \times 10^{5}$ & 34.038 & 0.0160 \\
Bin 2 & $4.368 \times 10^{3}$ & $2.348 \times 10^{3} - 1.452 \times 10^{4}$ & 73 & $3.189 \times 10^{5}$ & 34.176 & 0.119 \\
Bin 3 & $3.522 \times 10^{4}$ & $1.881 \times 10^{4} - 8.567 \times 10^{4}$ & 9 & $3.170 \times 10^{5}$ & 33.900 & 0.0786 \\

\hline
\label{tab:bin}
\end{tabular}
\footnotesize{$ ^a$Propagated in log$_{10}$ scale.}
\end{table*}

In order to compare the observational uIMFs to theoretical ones, we generate models using STARBURST99 (SB99; \citealt{Leitherer1999}) with a standard Kroupa IMF up to m$_{\mathrm{max}}$=120~M$_{\odot}$, Padova stellar evolution tracks with AGB stars and metallicity $Z = 0.004$. The choice of metallicity for the models is consistent with the metallicity of the sample galaxies (Table~\ref{tab:object}). The data points in Figure~\ref{fig:ratio} are compared with cluster models with both fixed $m_{\mathrm{max}}$ but different ages (left panel) and different $m_{\mathrm{max}}$ averaged over the age range  1--4~Myr (right panel). After distributing the clusters into 3 mass bins, the total equivalent masses in each bin are $3.186 \times 10^5 \: M_{\odot}$, $3.189 \times 10^5 \: M_{\odot}$ and $3.170 \times 10^5 \: M_{\odot}$ for the mass bins centered at $1.087 \times 10^3 \: M_{\odot}$, $4.368 \times 10^3 \: M_{\odot}$ and $3.522 \times 10^4 \: M_{\odot}$, respectively. The mass range of each bin is $5.172 \times 10^{2} \: M_{\odot} \, \mbox{--} \, 2.317 \times 10^{3} \:  M_{\odot}$, $2.348 \times 10^{3} \: M_{\odot} \, \mbox{--} \, 1.452 \times 10^{4} \: M_{\odot}$ and $1.881 \times 10^{4} \: M_{\odot} \, \mbox{--} \, 8.567 \times 10^{4} \: M_{\odot}$, respectively. More details on the binned data for each galaxy and for the combined sample are given in Table~\ref{tab:bin}. To test the impact of upper limits on our binned values, we replace all upper limits with  zero luminosity in H$\alpha$ ($L_{\mathrm{H} \alpha} = 0$). We find that that impact is negligible due to the small contribution ($\lesssim 1.6 \%$) of the upper limits to the mean value of $L_{\mathrm{H} \alpha} / M_{\mathrm{cl}}$ in Eq.(\ref{eq:ratio_avg}) (Figure~\ref{fig:individual}).

The distribution of $\langle L_{\mathrm{H} \alpha} / M_{\mathrm{cl}} \rangle$ for the combined sample shows a plateau as a function of mass rather than following the increasing trend for increasing mass described by the $m_{\mathrm{max}} \mbox{--} M_{\mathrm{ecl}}$ relation \citep{Weidner2010}. Taking $\langle L_{\mathrm{H} \alpha} / M_{\mathrm{cl}} \rangle \sim 10^{34} \,$erg$\cdot$s$^{-1}$$\cdot$$M_{\odot}^{-1}$ as our fiducial value at high cluster masses, the expected value at $M_{\mathrm{cl}} = 10^{3} \: M_{\odot}$ ($M_{\mathrm{ecl}} \sim 2 \times 10^{3} \: M_{\odot}$) should be about 0.5~dex lower according to the IGIMF models. We don't observe such a decrease, which is several standard deviations below our fiducial value at that mass (Figure~\ref{fig:ratio}). Furthermore, the $\langle L_{\mathrm{H} \alpha} / M_{\mathrm{cl}} \rangle$ values in the highest mass bins,  $M_{\mathrm{cl}} \sim 10^{4.5} \: M_{\odot}$, are lower than those at lower mass, similarly to what observed in NGC 4214 \citep{Andrews2013}.

The most likely reason for this decrease is that massive star clusters statistically host massive stars more frequently than lower-mass clusters, implying that feedback from those massive stars can effectively expel gas from the cluster location, an effect magnified in clusters residing in low H I density regions \citep{Pellegrini2012}. Gas expulsion implies that there is less gas nearby to produce H$\alpha$. This is observed in the most massive cluster in NGC 4214, where the presence of P-Cygni profiles for UV lines like Si IV ($\lambda\sim$1400~\AA) and C IV ($\lambda\sim$1550~\AA) indicates young ages, $\sim$4~Myr \citep{Leitherer1996, Leitherer2002}, but the cluster occupies a cavity in ionized gas \citep{mackenty2000}.

The horizontal dotted lines in the left panel (Figure~\ref{fig:ratio}) show the theoretical $L_{\mathrm{H} \alpha} / M_{\mathrm{cl}}$ as a function of cluster age. We expand the range of ages to 6 Myr which is broader than in our sample selection, because we include the uncertainties in the age determinations of the clusters, estimated to be 1--2~Myr \citep{Adamo2017}. There is almost no difference in the theoretical values of the ratios among 1--3~Myr located at $L_{\mathrm{H} \alpha} / M_{\mathrm{cl}} \sim 34.8$. The $L_{\mathrm{H} \alpha} / M_{\mathrm{cl}}$ becomes lower when the age of the cluster increases to 6 Myr ($L_{\mathrm{H} \alpha} / M_{\mathrm{cl}} \sim 33.7$). Nearly all data points are above $L_{\mathrm{H} \alpha} / M_{\mathrm{cl}}$ at 6 Myr except those of the highest mass bins in NGC 1313 and NGC 4395 where most of the sources undetected in H$\alpha$ are located in shell-like or diffused H II regions, possibly due to massive stellar feedback. 

The right panel in Figure~\ref{fig:ratio} shows the comparison of the observed $\langle L_{\mathrm{H} \alpha} / M_{\mathrm{cl}} \rangle$ and the theoretical values of $L_{\mathrm{H} \alpha} / M_{\mathrm{cl}}$ as a function of $m_{\mathrm{max}}$. The theoretical uIMFs are calculated using SB99 with different $m_{\mathrm{max}}$ values from $20 \: M_{\odot}$ to $120 \: M_{\odot}$ in $10 \: M_{\odot}$ steps. The H$\alpha$ luminosities are averaged within the 1--4 Myr age range for each $m_{\mathrm{max}}$ bin. The uIMFs of all data points are distributed below $m_{\mathrm{max}} \sim \: 70 \: M_{\odot}$ at $m_{\mathrm{max}} \sim 30 \mbox{--} 40 \: M_{\odot}$. As already remarked, this may point to the ages of our sample clusters being somewhat older than our assumed $\sim$4~Myr, and consistent with ages as old as $\sim$6~Myr.

Although the distribution of $\langle L_{\mathrm{H} \alpha} / M_{\mathrm{cl}} \rangle$ is relatively flat as a function of cluster mass in the mass range $\sim 500 \mbox{--} 50,000 \: M_{\odot}$ (Figure~\ref{fig:ratio}), we find that  the ionizing photon flux is significantly lower than expected for young ($\leq 4$~Myr) star clusters populated with a universal Kroupa IMF \citep{Kroupa2001} with upper mass cut off heavier than $100 \: M_{\odot}$. In other words, the combined sample data (gray squares) are located in correspondence of the $\langle L_{\mathrm{H} \alpha} / M_{\mathrm{cl}} \rangle$ values expected for populations with ages between 5~Myr and 6~Myr. In addition to the uncertainties in the ages, as already discussed above, we consider here ionizing photon leakage out of the star clusters. Although we attempt to recover as much of the $L_{\mathrm{H} \alpha}$ as possible, we cannot measure ionizing photons leakage directly. The escape fraction of ionizing photons from H II regions is estimated to be around 40--60\% by previous studies of diffuse ionized gas (DIG) in nearby galaxies \citep{Oey2007, Pellegrini2012, Belfiore2022}. We have some indirect evidence of this process from the difference in the average $L_{\mathrm{H} \alpha} / M_{\mathrm{cl}}$ values for NGC 1313 and NGC 4449. The average values for the clusters in NGC 4449 are a factor of $\sim$~5 systematically higher than those of NGC 1313, for all mass bins. NGC 4449 is a higher density galaxy than NGC 1313; comparing their SFR surface densities as a proxy for gas density, previous studies show $\Sigma_{\mathrm{SFR}} \sim 0.04 \: M_{\odot} \cdot \mathrm{yr}^{-1} \cdot \mathrm{kpc}^{-2}$ for NGC 4449 \citep{Annibali2011} and $\Sigma_{\mathrm{SFR}} \sim 0.01 \: M_{\odot} \cdot \mathrm{yr}^{-1} \cdot \mathrm{kpc}^{-2}$ for NGC 1313 \citep{Silva-Villa2011, Messa2021}. Thus, to the extent that lower density regions are also more porous \citep{Pellegrini2012}, our data support significant ionizing photon leakage out of the H II regions surrounding the star clusters in NGC 1313. This finding is in line with other results, indicating that high-SFR regions, like, e.g., the central starburst in the spiral M~83, may be less porous than low-SFR regions and contain clusters that do not tend to leak ionizing photons \citep{DellaBruna2022}.

In the binned mass plot (Figures~\ref{fig:ratio}), NGC 7793 is the only galaxy with an obvious trend of increasing $\langle L_{\mathrm{H} \alpha} / M_{\mathrm{cl}} \rangle$ for increasing mass, which may support the $m_{\mathrm{max}} \mbox{--} M_{\mathrm{ecl}}$ model. However, the other four galaxies do not display a comparable trend, and NGC 7793 does not contain clusters with mass $>1.2 \times 10^{4} \: M_{\odot}$, which leaves any inference about trends inconclusive. Furthermore, in the lowest mass bin in NGC 7793, 75\% of the clusters have $\lesssim 10^{3} M_{\mathrm{cl}}$ and $L_{\mathrm{H} \alpha} / M_{\mathrm{cl}} < 10^{33.4}$. This characteristic produces a mass bin with a value of  $\langle L_{\mathrm{H} \alpha} / M_{\mathrm{cl}} \rangle$ that is much lower for NGC 7793 than for other galaxies. As a counterpoint, only 23\% of the clusters that make the NGC 4449’s lowest-mass bin have $\lesssim 10^{3} M_{\mathrm{cl}}$, and this flattens considerably the trend of $\langle L_{\mathrm{H} \alpha} / M_{\mathrm{cl}} \rangle$ with cluster mass for this galaxy.

Finally, irrespective of the observed trend, there are numerous low-mass clusters, $\sim 10^3 \: M_{\odot}$ and lower, in all the sample galaxies but NGC 1705, that have significant emission in H$\alpha$ (Figure~\ref{fig:individual}), even stronger emission than the sources at masses $\gtrsim 5000 \: M_{\odot}$. The presence of H$\alpha$ emitting low-mass clusters indicates that stochasticity is a major contributor to the ionized gas emission of star clusters, further supporting our results.

\section{Summary and Conclusion} \label{sec:conclusion}

We have investigated the possibility of a non-universal upper-end for the stellar IMF (uIMF) using catalogs of young star clusters. Star clusters represent prime targets for this test due to the quasi-single-age nature of their stellar populations. 

We choose five nearby galaxies within $\sim 5$ Mpc with a broad range of SFRs as calculated from the dust-corrected UV ($0.1 \mbox{--} 1.2 \: M_{\odot} \: \mathrm{yr}^{-1}$) from the LEGUS data \citep{Calzetti2015a}. The cluster catalogs of these galaxies \citep{Adamo2017} are used to test variations of uIMF and we limit our analysis to young stellar clusters (YSCs) and compact associations with ages within 1--4~Myr and masses over $500 \: M_{\odot}$ (as determined from SED fits) to mitigate uncertainties due to the evolution of massive stars out of the main sequence in older clusters. We include sources without H$\alpha$ emission to prevent our results from being affected by biases, due to the stochastic nature of the IMF sampling, which disfavors presence of massive stars in low-mass clusters. 

Our analysis includes photometry of the H$\alpha$ emission in correspondence of all young star clusters identified in the five galaxies. The H$\alpha$ emission is obtained from imaging in the HST H$\alpha$+[N II] (F657N or F658N) filters, after stellar continuum subtraction using the interpolated stellar emission between V-short and I, as well as removal of the [N II] contamination in the F657N/F658N filters. We apply aperture corrections to account for the ionized gas outside of the photometric apertures by using growth curves derived from isolated H$\alpha$-emitting sources. 

Our results, shown in Figure~\ref{fig:ratio}, are consistent with those of \citet{Andrews2013, Andrews2014}, in that the mass-normalized ionizing photon flux is basically constant with cluster mass. This implies that the IMF from which the cluster populations are drawn is universal. The $\langle L_{\mathrm{H} \alpha} / M_{\mathrm{cl}} \rangle$ values in Figure~\ref{fig:ratio} do not strongly follow the trend expected in the $m_{\mathrm{max}} \mbox{--} M_{\mathrm{ecl}}$ relation of \citet{Weidner2010}. The three mass bins of the combined clusters are centered at $1 \times 10^3 \: M_{\odot}$, $4.4 \times 10^3 \: M_{\odot}$ and $3.5 \times 10^4 \: M_{\odot}$, shown as gray squares in Figure~\ref{fig:ratio}, and their $\langle L_{\mathrm{H} \alpha} / M_{\mathrm{cl}} \rangle$ values mark a relatively constant trend, around $\sim 30 \mbox{--} 40 \: M_{\odot}$ for the most massive stars. This value  can be explained if our sample, which is selected to include clusters up to $\sim$~4~Myr, actually includes clusters potentially as old as $\sim$~6~Myr, due to uncertainties in the age determinations, as well as significant ionizing photon leakage out of the H II regions. Leakage can lower the measured H$\alpha$ luminosity by about a factor of 2, on average \citep{Oey2007, Pellegrini2012, Belfiore2022}.

Looking at individual galaxies, only NGC 7793 shows a trend consistent with the $m_{\mathrm{max}} \mbox{--} M_{\mathrm{ecl}}$ relation  (Figures~\ref{fig:ratio}). However, this galaxy lacks clusters more massive than $1.2 \times 10^{4} \: M_{\odot}$, which leaves the issue unsettled. In addition, several low-mass clusters in NGC 7793, down to $\sim 10^{3} \: M_{\odot}$, show significant H$\alpha$ emission (Figure~\ref{fig:individual}), implying the low-mass clusters host massive stars. 

The $\langle L_{\mathrm{H} \alpha} / M_{\mathrm{cl}} \rangle$ values of the heaviest mass bin at $\sim 10^{4.5} \: M_{\odot}$ have lower values than those of other mass bins both for individual galaxies and for the combined sample. This finding had been reported in \citet{Andrews2013, Andrews2014} and is likely caused by feedback effects from the faster dynamical evolution in higher mass clusters.

We have confirmed that the uIMF is consistent with a universal trend, and deviations are due to stochasticity, in good agreement with previous studies \citep{Calzetti2010, Andrews2013, Andrews2014}. More cluster samples, especially from galaxies spanning a larger range of metallicities and SFRs than those explored here, would be desirable to provide a definite test of the dependence of the maximum mass stars on the cluster masses where they are embedded.

Furthermore, infrared observations would allow us to obtain a census of the clusters that are still deeply embedded in their natal clouds, and are missing from our accounting. These tend to be the youngest among our clusters, typically with ages $\leq 3$~Myr \citep{Messa2021}. These embedded clusters are the most promising for sampling the youngest stars and the unexpanded gas regions they ionize. Future observations with the James Webb Space Telescope (JWST) are likely to provide such samples of embedded clusters even in more distant galaxies then  5 Mpc, thanks to Webb's better resolution in the infrared than the Hubble Space Telescope.

\begin{acknowledgments}

We thank the LEGUS Survey team for providing us with the high-quality observational data and catalogs to support this research.
\end{acknowledgments}

\facilities{HST(ACS and WFC3)}
\software{SExtractor \citep{Bertin1996}, Astropy \citep{Astropy2022}}

\vspace{5mm}

%%\FloatBarrier
\bibliography{references}{}

\begin{thebibliography}{}
\expandafter\ifx\csname natexlab\endcsname\relax\def\natexlab#1{#1}\fi
\providecommand{\url}[1]{\href{#1}{#1}}
\providecommand{\dodoi}[1]{doi:~\href{http://doi.org/#1}{\nolinkurl{#1}}}
\providecommand{\doeprint}[1]{\href{http://ascl.net/#1}{\nolinkurl{http://ascl.net/#1}}}
\providecommand{\doarXiv}[1]{\href{https://arxiv.org/abs/#1}{\nolinkurl{https://arxiv.org/abs/#1}}}

\bibitem[{Adamo {et~al.}(2017)Adamo, Ryon, Messa, Kim, Grasha, Cook, Calzetti,
  Lee, Whitmore, Elmegreen, Ubeda, Smith, Bright, Runnholm, Andrews, Fumagalli,
  Gouliermis, Kahre, Nair, Thilker, Walterbos, Wofford, Aloisi, Ashworth,
  Brown, Chandar, Christian, Cignoni, Clayton, Dale, de~Mink, Dobbs, Elmegreen,
  Evans, {Gallagher III}, Grebel, Herrero, Hunter, Johnson, Kennicutt,
  Krumholz, Lennon, Levay, Martin, Nota, {\"{O}}stlin, Pellerin, Prieto, Regan,
  Sabbi, Sacchi, Schaerer, Schiminovich, Shabani, Tosi, {Van Dyk}, \&
  Zackrisson}]{Adamo2017}
Adamo, A., Ryon, J.~E., Messa, M., {et~al.} 2017, Astrophys. J., 841, 131,
  \dodoi{10.3847/1538-4357/aa7132}

\bibitem[{Andrews {et~al.}(2013)Andrews, Calzetti, Chandar, Lee, Elmegreen,
  Kennicutt, Whitmore, Kissel, {Da Silva}, Krumholz, O'Connell, Dopita, Frogel,
  \& Kim}]{Andrews2013}
Andrews, J.~E., Calzetti, D., Chandar, R., {et~al.} 2013, Astrophys. J., 767,
  51, \dodoi{10.1088/0004-637X/767/1/51}

\bibitem[{Andrews {et~al.}(2014)Andrews, Calzetti, Chandar, Elmegreen,
  Kennicutt, Kim, Krumholz, Lee, McElwee, O'Connell, \& Whitmore}]{Andrews2014}
---. 2014, Astrophys. J., 793, 4, \dodoi{10.1088/0004-637X/793/1/4}

\bibitem[{Annibali {et~al.}(2011)Annibali, Tosi, Aloisi, \& {Van Der
  Marel}}]{Annibali2011}
Annibali, F., Tosi, M., Aloisi, A., \& {Van Der Marel}, R.~P. 2011, Astron. J.,
  142, \dodoi{10.1088/0004-6256/142/4/129}

\bibitem[{Ascenso {et~al.}(2009)Ascenso, Alves, \& Lago}]{Ascenso2009}
Ascenso, J., Alves, J., \& Lago, M.~T. 2009, Astron. Astrophys., 495, 147,
  \dodoi{10.1051/0004-6361:200809886}

\bibitem[{Bastian {et~al.}(2010)Bastian, Covey, \& Meyer}]{Bastian2010}
Bastian, N., Covey, K.~R., \& Meyer, M.~R. 2010, Annu. Rev. Astron. Astrophys.,
  48, 339, \dodoi{10.1146/annurev-astro-082708-101642}

\bibitem[{Belfiore {et~al.}(2022)Belfiore, Santoro, Groves, Schinnerer,
  Kreckel, Glover, Klessen, Emsellem, Blanc, Congiu, Barnes, Boquien, Chevance,
  Dale, {Diederik Kruijssen}, Leroy, Pan, Pessa, Schruba, \&
  Williams}]{Belfiore2022}
Belfiore, F., Santoro, F., Groves, B., {et~al.} 2022, Astron. Astrophys., 659,
  \dodoi{10.1051/0004-6361/202141859}

\bibitem[{Bertin \& Arnouts(1996)}]{Bertin1996}
Bertin, E., \& Arnouts, S. 1996, Astron. Astrophys., 117, 393.
\newblock \doarXiv{arXiv:1011.1669v3}

\bibitem[{Brands {et~al.}(2022)Brands, {De Koter}, Bestenlehner, Crowther,
  Sundqvist, Puls, Caballero-Nieves, Abdul-Masih, Driessen, Garc{\'{i}}a, Geen,
  Gr{\"{a}}fener, Hawcroft, Kaper, Keszthelyi, Langer, Sana, Schneider, Shenar,
  \& Vink}]{Brands2022}
Brands, S.~A., {De Koter}, A., Bestenlehner, J.~M., {et~al.} 2022, Astron.
  Astrophys., 663, \dodoi{10.1051/0004-6361/202142742}

\bibitem[{Calzetti {et~al.}(2000)Calzetti, Armus, Bohlin, Kinney, Koornneef, \&
  Storchi‐Bergmann}]{Calzetti2000}
Calzetti, D., Armus, L., Bohlin, R.~C., {et~al.} 2000, Astrophys. J., 533, 682,
  \dodoi{10.1086/308692}

\bibitem[{Calzetti {et~al.}(2010)Calzetti, Chandar, Lee, Elmegreen, Kennicutt,
  \& Whitmore}]{Calzetti2010}
Calzetti, D., Chandar, R., Lee, J.~C., {et~al.} 2010, Astrophys. J. Lett., 719,
  158, \dodoi{10.1088/2041-8205/719/2/L158}

\bibitem[{Calzetti {et~al.}(1994)Calzetti, Kinney, \&
  Storchi-Bergmann}]{Calzetti1994}
Calzetti, D., Kinney, L.~A., \& Storchi-Bergmann, T. 1994, Astrophys. J., 429,
  582

\bibitem[{Calzetti {et~al.}(2015{\natexlab{a}})Calzetti, Lee, Sabbi, Adamo,
  Smith, Andrews, Ubeda, Bright, Thilker, Aloisi, Brown, Chandar, Christian,
  Cignoni, Clayton, {Da Silva}, {De Mink}, Dobbs, Elmegreen, Elmegreen, Evans,
  Fumagalli, Gallagher, Gouliermis, Grebel, Herrero, Hunter, Johnson,
  Kennicutt, Kim, Krumholz, Lennon, Levay, Martin, Nair, Nota, {\"{O}}stlin,
  Pellerin, Prieto, Regan, Ryon, Schaerer, Schiminovich, Tosi, {Van Dyk},
  Walterbos, Whitmore, \& Wofford}]{Calzetti2015a}
Calzetti, D., Lee, J.~C., Sabbi, E., {et~al.} 2015{\natexlab{a}}, Astron. J.,
  149, 51, \dodoi{10.1088/0004-6256/149/2/51}

\bibitem[{Calzetti {et~al.}(2015{\natexlab{b}})Calzetti, Johnson, Adamo,
  Gallagher, Andrews, Smith, Clayton, Lee, Sabbi, Ubeda, Kim, Ryon, Thilker,
  Bright, Zackrisson, Kennicutt, Mink, Whitmore, Aloisi, Chandar, Cignoni,
  Cook, Dale, Elmegreen, Elmegreen, Evans, Fumagalli, Gouliermis, Grasha,
  Grebel, Krumholz, Walterbos, Wofford, Brown, Christian, Dobbs, Herrero,
  Kahre, Messa, Nair, Nota, {\"{O}}stlin, Pellerin, Sacchi, Schaerer, \&
  Tosi}]{Calzetti2015b}
Calzetti, D., Johnson, K.~E., Adamo, A., {et~al.} 2015{\natexlab{b}},
  Astrophys. J., 811, 75, \dodoi{10.1088/0004-637X/811/2/75}

\bibitem[{Cappellari {et~al.}(2012)Cappellari, McDermid, Alatalo, Blitz, Bois,
  Bournaud, Bureau, Crocker, Davies, Davis, {De Zeeuw}, Duc, Emsellem,
  Khochfar, Krajnovi{\"{a}}, Kuntschner, Lablanche, Morganti, Naab, Oosterloo,
  Sarzi, Scott, Serra, Weijmans, \& Young}]{Cappellari2012}
Cappellari, M., McDermid, R.~M., Alatalo, K., {et~al.} 2012, Nature, 484, 485,
  \dodoi{10.1038/nature10972}

\bibitem[{Cervino {et~al.}(2002)Cervino, Valls-Gabaud, Luridiana, \&
  Mas-Hesse}]{Cervino2002}
Cervino, M., Valls-Gabaud, D., Luridiana, V., \& Mas-Hesse, J.~M. 2002, Astron.
  Astrophys., 381, 51, \dodoi{10.1051/0004-6361}

\bibitem[{Chabrier(2003)}]{Chabrier2003}
Chabrier, G. 2003, Publ. Astron. Soc. Pacific, 115, 763, \dodoi{10.1086/376392}

\bibitem[{Cook {et~al.}(2019)Cook, Lee, Adamo, Kim, Chandar, Whitmore, Mok,
  Ryon, Dale, Calzetti, Andrews, Aloisi, Ashworth, Bright, Brown, Christian,
  Cignoni, Clayton, {Da Silva}, {De Mink}, Dobbs, Elmegreen, Elmegreen, Evans,
  Fumagalli, Gallagher, Gouliermis, Grasha, Grebel, Herrero, Hunter, Jensen,
  Johnson, Kahre, Kennicutt, Krumholz, Lee, Lennon, Linden, Martin, Messa,
  Nair, Nota, {\"{O}}stlin, Parziale, Pellerin, Regan, Sabbi, Sacchi, Schaerer,
  Schiminovich, Shabani, Slane, Small, Smith, Smith, Taibi, Thilker, {De La
  Torre}, Tosi, Turner, Ubeda, {Van Dyk}, Walterbos, \& Wofford}]{Cook2019}
Cook, D.~O., Lee, J.~C., Adamo, A., {et~al.} 2019, Mon. Not. R. Astron. Soc.,
  484, 4897, \dodoi{10.1093/mnras/stz331}

\bibitem[{Corbelli {et~al.}(2009)Corbelli, Verley, Elmegreen, \&
  Giovanardi}]{Corbelli2009}
Corbelli, E., Verley, S., Elmegreen, B.~G., \& Giovanardi, C. 2009, Astron.
  Astrophys., 495, 479, \dodoi{10.1051/0004-6361:200811086}

\bibitem[{Crowther {et~al.}(2010)Crowther, Schnurr, Hirschi, Yusof, Parker,
  Goodwin, \& Kassim}]{Crowther2010}
Crowther, P.~A., Schnurr, O., Hirschi, R., {et~al.} 2010, Mon. Not. R. Astron.
  Soc., 408, 731, \dodoi{10.1111/j.1365-2966.2010.17167.x}

\bibitem[{Dabringhausen {et~al.}(2008)Dabringhausen, Hilker, \&
  Kroupa}]{Dabringhausen2008}
Dabringhausen, J., Hilker, M., \& Kroupa, P. 2008, Mon. Not. R. Astron. Soc.,
  386, 864, \dodoi{10.1111/j.1365-2966.2008.13065.x}

\bibitem[{{Della Bruna} {et~al.}(2022){Della Bruna}, Adamo, Mcleod, Smith,
  Savard, Robert, Sun, Amram, Bik, Blair, Long, Renaud, Walterbos, \&
  Usher}]{DellaBruna2022}
{Della Bruna}, L., Adamo, A., Mcleod, A.~F., {et~al.} 2022, Astron. Astrophys.,
  666, A29, \dodoi{10.1051/0004-6361/202243395}

\bibitem[{Elmegreen(1983)}]{Elmegreen1983}
Elmegreen, B.~G. 1983, Mon. Not. R. Astron. Soc., 203, 1011

\bibitem[{Elmegreen(2000)}]{Elmegreen2000}
---. 2000, ESA SP, 445, 265

\bibitem[{Elmegreen(2006)}]{Elmegreen2006a}
---. 2006, Astrophys. J., 648, 572, \dodoi{10.1086/505785}

\bibitem[{Elmegreen \& Scalo(2006)}]{Elmegreen2006}
Elmegreen, B.~G., \& Scalo, J. 2006, Astrophys. J., 636, 149,
  \dodoi{10.1086/497889}

\bibitem[{Fardal {et~al.}(2007)Fardal, Katz, Weinberg, \&
  Dav{\'{e}}}]{Fardal2007}
Fardal, M.~A., Katz, N., Weinberg, D.~H., \& Dav{\'{e}}, R. 2007, Mon. Not. R.
  Astron. Soc., 379, 985, \dodoi{10.1111/j.1365-2966.2007.11522.x}

\bibitem[{Fumagalli {et~al.}(2011)Fumagalli, {Da Silva}, \&
  Krumholz}]{Fumagalli2011}
Fumagalli, M., {Da Silva}, R.~L., \& Krumholz, M.~R. 2011, Astrophys. J. Lett.,
  741, 1, \dodoi{10.1088/2041-8205/741/2/L26}

\bibitem[{Geha {et~al.}(2013)Geha, Brown, Tumlinson, Kalirai, Simon, Kirby,
  Vandenberg, Mu{\~{n}}oz, Avila, Guhathakurta, \& Ferguson}]{Geha2013}
Geha, M., Brown, T.~M., Tumlinson, J., {et~al.} 2013, Astrophys. J., 771, 29,
  \dodoi{10.1088/0004-637X/771/1/29}

\bibitem[{Grasha {et~al.}(2015)Grasha, Calzetti, Adamo, Kim, Elmegreen,
  Gouliermis, Aloisi, Bright, Christian, Cignoni, Dale, Dobbs, Elmegreen,
  Fumagalli, Iii, Grebel, Johnson, Lee, Messa, Smith, Ryon, Thilker, Ubeda, \&
  Wofford}]{Grasha2015}
Grasha, K., Calzetti, D., Adamo, A., {et~al.} 2015, Astrophys. J., 815, 93,
  \dodoi{10.1088/0004-637X/815/2/93}

\bibitem[{Grasha {et~al.}(2017{\natexlab{a}})Grasha, Calzetti, Adamo, Kim,
  Elmegreen, Gouliermis, Dale, Fumagalli, Grebel, Johnson, Kahre, Kennicutt,
  Messa, Pellerin, Ryon, Smith, Shabani, Thilker, \& Ubeda}]{Grasha2017a}
---. 2017{\natexlab{a}}, Astrophys. J., 840, 113,
  \dodoi{10.3847/1538-4357/aa6f15}

\bibitem[{Grasha {et~al.}(2017{\natexlab{b}})Grasha, Elmegreen, Calzetti,
  Adamo, Aloisi, Bright, Cook, Dale, Fumagalli, {Gallagher III}, Gouliermis,
  Grebel, Kahre, Kim, Krumholz, Lee, Messa, Ryon, \& Ubeda}]{Grasha2017}
Grasha, K., Elmegreen, B.~G., Calzetti, D., {et~al.} 2017{\natexlab{b}},
  Astrophys. J., 842, 25, \dodoi{10.3847/1538-4357/aa740b}

\bibitem[{Grasha {et~al.}(2018)Grasha, Calzetti, Bittle, Johnson, {Donovan
  Meyer}, Kennicutt, Elmegreen, Adamo, Krumholz, Fumagalli, Grebel, Gouliermis,
  Cook, Gallagher, Aloisi, Dale, Linden, Sacchi, Thilker, Walterbos, Messa,
  Wofford, \& Smith}]{Grasha2018}
Grasha, K., Calzetti, D., Bittle, L., {et~al.} 2018, Mon. Not. R. Astron. Soc.,
  481, 1016, \dodoi{10.1093/MNRAS/STY2154}

\bibitem[{Grasha {et~al.}(2019)Grasha, Calzetti, Adamo, Kennicutt, Elmegreen,
  Messa, Dale, Fedorenko, Mahadevan, Grebel, Fumagalli, Kim, Dobbs, Gouliermis,
  Ashworth, Gallagher, Smith, Tosi, Whitmore, Schinnerer, Colombo, Hughes,
  Leroy, \& Meidt}]{Grasha2019}
Grasha, K., Calzetti, D., Adamo, A., {et~al.} 2019, Mon. Not. R. Astron. Soc.,
  483, 4707, \dodoi{10.1093/mnras/sty3424}

\bibitem[{Hermanowicz {et~al.}(2013)Hermanowicz, Kennicutt, \&
  Eldridge}]{Hermanowicz2013}
Hermanowicz, M.~T., Kennicutt, R.~C., \& Eldridge, J.~J. 2013, Mon. Not. R.
  Astron. Soc., 432, 3097, \dodoi{10.1093/mnras/stt665}

\bibitem[{Jeř{\'{a}}bkov{\'{a}} {et~al.}(2018)Jeř{\'{a}}bkov{\'{a}}, {Hasani
  Zonoozi}, Kroupa, Beccari, Yan, Vazdekis, \& Zhang}]{Jerabkova2018}
Jeř{\'{a}}bkov{\'{a}}, T., {Hasani Zonoozi}, A., Kroupa, P., {et~al.} 2018,
  Astron. Astrophys., 620, 1, \dodoi{10.1051/0004-6361/201833055}

\bibitem[{Kalari {et~al.}(2022)Kalari, Horch, Salinas, Vink, Andersen,
  Bestenlehner, \& Rubio}]{Kalari2022}
Kalari, V.~M., Horch, E.~P., Salinas, R., {et~al.} 2022, Astrophys. J., 935,
  162, \dodoi{10.3847/1538-4357/ac8424}

\bibitem[{{Kennicutt, Jr.}(1998)}]{Kennicutt1998}
{Kennicutt, Jr.}, R.~C. 1998, Astrophys. J., 498, 541, \dodoi{10.1086/305588}

\bibitem[{{Kennicutt, Jr.} {et~al.}(2008){Kennicutt, Jr.}, Lee, {Funes, S. J.},
  Sakai, \& Akiyama}]{KennicuttJr.2008}
{Kennicutt, Jr.}, R.~C., Lee, J.~C., {Funes, S. J.}, J.~G., Sakai, S., \&
  Akiyama, S. 2008, Astrophys. J. Suppl. Ser., 178, 247, \dodoi{10.1086/590058}

\bibitem[{Kroupa(2001)}]{Kroupa2001}
Kroupa, P. 2001, Mon. Not. R. Astron. Soc., 322, 231,
  \dodoi{10.1046/j.1365-8711.2001.04022.x}

\bibitem[{Kroupa \& Boily(2002)}]{KroupaBoily2002}
Kroupa, P., \& Boily, C.~M. 2002, Mon. Not. R. Astron. Soc., 336, 1188,
  \dodoi{10.1046/j.1365-8711.2002.05848.x}

\bibitem[{Kroupa \& Weidner(2003)}]{Kroupa2003}
Kroupa, P., \& Weidner, C. 2003, Astrophys. J., 598, 1076,
  \dodoi{10.1086/379105}

\bibitem[{Kroupa {et~al.}(2012)Kroupa, Weidner, Pflamm-Altenburg, Thies,
  Dabringhausen, Marks, \& Maschberger}]{Kroupa2012}
Kroupa, P., Weidner, C., Pflamm-Altenburg, J., {et~al.} 2012, Planets, Stars
  Stellar Syst. Vol. 5 Galact. Struct. Stellar Popul., 5, 115,
  \dodoi{10.1007/978-94-007-5612-0_4}

\bibitem[{Krumholz {et~al.}(2019)Krumholz, McKee, \&
  Bland-Hawthorn}]{Krumholz2019}
Krumholz, M.~R., McKee, C.~F., \& Bland-Hawthorn, J. 2019, Annu. Rev. Astron.
  Astrophys., 57, 227, \dodoi{10.1146/annurev-astro-091918-104430}

\bibitem[{Lada \& Lada(2003)}]{Lada2003}
Lada, C.~J., \& Lada, E.~A. 2003, Annu. Rev. Astron. Astrophys., 41, 57,
  \dodoi{10.1146/annurev.astro.41.011802.094844}

\bibitem[{Larson(1982)}]{Larson1982}
Larson, R.~B. 1982, Mon. Not. R. Astron. Soc., 200, 159.
\newblock \url{http://eprints.uanl.mx/5481/1/1020149995.PDF}

\bibitem[{Leitherer {et~al.}(2002)Leitherer, Calzetti, \&
  Martins}]{Leitherer2002}
Leitherer, C., Calzetti, D., \& Martins, L.~P. 2002, Astrophys. J., 574, 114,
  \dodoi{10.1086/340902}

\bibitem[{Leitherer {et~al.}(1999)Leitherer, Schaerer, Goldader, Gonza,
  Delgado, Kune, Devost, \& Heckman}]{Leitherer1999}
Leitherer, C., Schaerer, D., Goldader, J.~D., {et~al.} 1999, ApJS, 1, 3

\bibitem[{Leitherer {et~al.}(1996)Leitherer, Vacca, Conti, Filippenko, Robert,
  \& Sargent}]{Leitherer1996}
Leitherer, C., Vacca, W.~D., Conti, P.~S., {et~al.} 1996, Astrophys. J., 465,
  717.
\newblock \doarXiv{تم}

\bibitem[{MacKenty {et~al.}(2000)MacKenty, Maiz-Apellaniz, Pickens, Norman, \&
  Walborn}]{mackenty2000}
MacKenty, J.~W., Maiz-Apellaniz, J.~S., Pickens, C.~E., Norman, C.~A., \&
  Walborn, N.~R. 2000, Astron. J., 2, 894

\bibitem[{Ma{\'{i}}z-Apellniz(2008)}]{MaizApellniz2008}
Ma{\'{i}}z-Apellniz, J. 2008, Astrophys. J., 677, 1278,
  \dodoi{10.1088/0004-637X/699/2/1938}

\bibitem[{Marques-Chaves {et~al.}(2022)Marques-Chaves, Schaerer,
  Alvarez-Marquez, Verhamme, Ceverino, Chisholm, Colina, Dessauges-Zavadsky,
  Perez-Fournon, Saldana-Lopez, Upadhyaya, \& Vanzella}]{Marques-Chaves2022}
Marques-Chaves, R., Schaerer, D., Alvarez-Marquez, J., {et~al.} 2022, Mon. Not.
  R. Astron. Soc., 517, 2972.
\newblock \doarXiv{2210.02392}

\bibitem[{Massey(2003)}]{Massey2003}
Massey, P. 2003, Annu. Rev. Astron. Astrophys., 41, 15,
  \dodoi{10.1146/annurev.astro.41.071601.170033}

\bibitem[{Messa {et~al.}(2018{\natexlab{a}})Messa, Adamo, Ostlin, Calzetti,
  Grasha, Grebel, Shabani, Chandar, Dale, Dobbs, Elmegreen, Fumagalli,
  Gouliermis, Kim, Smith, Thilker, Tosi, Ubeda, Walterbos, Whitmore, Fedorenko,
  Mahadevan, Andrews, Bright, Cook, Kahre, Nair, Pellerin, Ryon, Ahmad, Beale,
  Brown, Clarkson, Guidarelli, Parziale, Turner, \& Weber}]{Messa2018a}
Messa, M., Adamo, A., Ostlin, G., {et~al.} 2018{\natexlab{a}}, Mon. Not. R.
  Astron. Soc., 473, 996, \dodoi{10.1093/mnras/stx2403}

\bibitem[{Messa {et~al.}(2018{\natexlab{b}})Messa, Adamo, Calzetti,
  Reina-Campos, Colombo, Schinnerer, Chandar, Dale, Gouliermis, Grasha, Grebel,
  Elmegreen, Fumagalli, Johnson, Kruijssen, {\"{O}}stlin, Shabani, Smith, \&
  Whitmore}]{Messa2018b}
Messa, M., Adamo, A., Calzetti, D., {et~al.} 2018{\natexlab{b}}, Mon. Not. R.
  Astron. Soc., 477, 1683, \dodoi{10.1093/mnras/sty577}

\bibitem[{Messa {et~al.}(2021)Messa, Calzetti, Adamo, Grasha, Johnson, Sabbi,
  Smith, Bajaj, Finn, \& Lin}]{Messa2021}
Messa, M., Calzetti, D., Adamo, A., {et~al.} 2021, Astrophys. J., 909, 121,
  \dodoi{10.3847/1538-4357/abe0b5}

\bibitem[{Miller \& Scalo(1979)}]{Miller1979}
Miller, G.~E., \& Scalo, J.~M. 1979, Astrophys. J. Suppl. Ser., 41, 513,
  \dodoi{10.1086/190629}

\bibitem[{Oey(2011)}]{Oey2011}
Oey, M.~S. 2011, Astrophys. J. Lett., 739, L46,
  \dodoi{10.1088/2041-8205/739/2/L46}

\bibitem[{Oey \& Clarke(1998)}]{Oey1998}
Oey, M.~S., \& Clarke, C.~J. 1998, Astron. J., 115, 1543.
\newblock \url{http://iopscience.iop.org/1538-3881/115/4/1543}

\bibitem[{Oey {et~al.}(2007)Oey, Meurer, Yelda, Furst, Caballero‐Nieves,
  Hanish, Levesque, Thilker, Walth, Bland‐Hawthorn, Dopita, Ferguson,
  Heckman, Doyle, Drinkwater, Freeman, {Kennicutt, Jr.}, Kilborn, Knezek,
  Koribalski, Meyer, Putman, Ryan‐Weber, Smith, Staveley‐Smith, Webster,
  Werk, \& Zwaan}]{Oey2007}
Oey, M.~S., Meurer, G.~R., Yelda, S., {et~al.} 2007, Astrophys. J., 661, 801,
  \dodoi{10.1086/517867}

\bibitem[{Orozco-Duarte {et~al.}(2022)Orozco-Duarte, Wofford,
  Vidal-Garc{\'{i}}a, Bruzual, Charlot, Krumholz, Hannon, Lee, Wofford,
  Fumagalli, Dale, Messa, Grebel, Smith, Grasha, \& Cook}]{Orozco-Duarte2022}
Orozco-Duarte, R., Wofford, A., Vidal-Garc{\'{i}}a, A., {et~al.} 2022, Mon.
  Not. R. Astron. Soc., 509, 522, \dodoi{10.1093/mnras/stab2988}

\bibitem[{Pellegrini {et~al.}(2012)Pellegrini, Oey, Winkler, Points, Smith,
  Jaskot, \& Zastrow}]{Pellegrini2012}
Pellegrini, E.~W., Oey, M.~S., Winkler, P.~F., {et~al.} 2012, Astrophys. J.,
  755, 40, \dodoi{10.1088/0004-637X/755/1/40}

\bibitem[{Pflamm-Altenburg {et~al.}(2009)Pflamm-Altenburg, Weidner, \&
  Kroupa}]{Pflamm-Altenburg2009}
Pflamm-Altenburg, J., Weidner, C., \& Kroupa, P. 2009, Mon. Not. R. Astron.
  Soc., 395, 394, \dodoi{10.1111/j.1365-2966.2009.14522.x}

\bibitem[{Rela{\~{n}}o {et~al.}(2012)Rela{\~{n}}o, Kennicutt, Eldridge, Lee, \&
  Verley}]{Relano2012}
Rela{\~{n}}o, M., Kennicutt, R.~C., Eldridge, J.~J., Lee, J.~C., \& Verley, S.
  2012, Mon. Not. R. Astron. Soc., 423, 2933,
  \dodoi{10.1111/j.1365-2966.2012.21107.x}

\bibitem[{Romano {et~al.}(2005)Romano, Karakas, Tosi, \&
  Matteucci}]{Romano2005}
Romano, D., Karakas, A.~I., Tosi, M., \& Matteucci, F. 2005, Astron.
  Astrophys., 430, 491, \dodoi{10.1051/0004-6361/201014483}

\bibitem[{Sabbi {et~al.}(2018)Sabbi, Calzetti, Ubeda, Adamo, Cignoni, Thilker,
  Aloisi, Elmegreen, Elmegreen, Gouliermis, Grebel, Messa, Smith, Tosi,
  Dolphin, Andrews, Ashworth, Bright, Brown, Chandar, Christian, Clayton, Cook,
  Dale, de~Mink, Dobbs, Evans, Fumagalli, Gallagher, Grasha, Herrero, Hunter,
  Johnson, Kahre, Kennicutt, Kim, Krumholz, Lee, Lennon, Martin, Nair, Nota,
  {\"{O}}stlin, Pellerin, Prieto, Regan, Ryon, Sacchi, Schaerer, Schiminovich,
  Shabani, {Van Dyk}, Walterbos, Whitmore, \& Wofford}]{Sabbi2018}
Sabbi, E., Calzetti, D., Ubeda, L., {et~al.} 2018, Astrophys. J. Suppl. Ser.,
  235, 23, \dodoi{10.3847/1538-4365/aaa8e5}

\bibitem[{Salpeter(1955)}]{Salpeter1955}
Salpeter, E.~E. 1955, Astrophys. J., 121, 161.
\newblock \url{https://ui.adsabs.harvard.edu/abs/1955ApJ...121..161S/abstract}

\bibitem[{Schneider {et~al.}(2018)Schneider, Sana, Evans, Bestenlehner, Castro,
  Fossati, Gr{\"{a}}fener, Langer, Ram{\'{i}}rez-Agudelo,
  Sab{\'{i}}n-Sanjuli{\'{a}}n, Sim{\'{o}}n-D{\'{i}}az, Tramper, Crowther,
  de~Koter, de~Mink, Dufton, Garcia, Gieles, H{\'{e}}nault-Brunet, Herrero,
  Izzard, Kalari, Lennon, Apell{\'{a}}niz, Markova, Najarro, Podsiadlowski,
  Puls, Taylor, van Loon, Vink, \& Norman}]{Schneider2018}
Schneider, F. R.~N., Sana, H., Evans, C.~J., {et~al.} 2018, Science (80-. ).,
  361, 69, \dodoi{10.1126/science.aat6506}

\bibitem[{Silva-Villa \& Larsen(2011)}]{Silva-Villa2011}
Silva-Villa, E., \& Larsen, S.~S. 2011, Astron. Astrophys., 529,
  \dodoi{10.1051/0004-6361/201016206}

\bibitem[{{The Astropy Collaboration} {et~al.}(2022){The Astropy
  Collaboration}, Price-Whelan, Lim, Earl, Starkman, Bradley, Shupe, Patil,
  Corrales, Brasseur, N{\"{o}}the, Donath, Tollerud, Morris, Ginsburg, Vaher,
  Weaver, Tocknell, Jamieson, van Kerkwijk, Robitaille, Merry, Bachetti,
  G{\"{u}}nther, Aldcroft, Alvarado-Montes, Archibald, B{\'{o}}di, Bapat,
  Barentsen, Baz{\'{a}}n, Biswas, Boquien, Burke, Cara, Cara, Conroy, Conseil,
  Craig, Cross, Cruz, D'Eugenio, Dencheva, Devillepoix, Dietrich, Eigenbrot,
  Erben, Ferreira, Foreman-Mackey, Fox, Freij, Garg, Geda, Glattly,
  Gondhalekar, Gordon, Grant, Greenfield, Groener, Guest, Gurovich, Handberg,
  Hart, Hatfield-Dodds, Homeier, Hosseinzadeh, Jenness, Jones, Joseph,
  Kalmbach, Karamehmetoglu, Ka{\l}uszy{\'{n}}ski, Kelley, Kern, Kerzendorf,
  Koch, Kulumani, Lee, Ly, Ma, MacBride, Maljaars, Muna, Murphy, Norman,
  O'Steen, Oman, Pacifici, Pascual, Pascual-Granado, Patil, Perren, Pickering,
  Rastogi, Roulston, Ryan, Rykoff, Sabater, Sakurikar, Salgado, Sanghi,
  Saunders, Savchenko, Schwardt, Seifert-Eckert, Shih, Jain, Shukla, Sick,
  Simpson, Singanamalla, Singer, Singhal, Sinha, Sipőcz, Spitler, Stansby,
  Streicher, {\v{S}}umak, Swinbank, Taranu, Tewary, Tremblay, de~Val-Borro,
  {Van Kooten}, Vasovi{\'{c}}, Verma, {de Miranda Cardoso}, Williams, Wilson,
  Winkel, Wood-Vasey, Xue, Yoachim, Zhang, \& Zonca}]{Astropy2022}
{The Astropy Collaboration}, Price-Whelan, A.~M., Lim, P.~L., {et~al.} 2022,
  Astrophys. J., 935, 167, \dodoi{10.3847/1538-4357/ac7c74}

\bibitem[{Tolstoy {et~al.}(2009)Tolstoy, Hill, \& Tosi}]{Tolstoy2009}
Tolstoy, E., Hill, V., \& Tosi, M. 2009, Annu. Rev. Astron. Astrophys., 47,
  371, \dodoi{10.1146/annurev-astro-082708-101650}

\bibitem[{{Van Dokkum} \& Conroy(2011)}]{VanDokkum2011}
{Van Dokkum}, P.~G., \& Conroy, C. 2011, Astrophys. J. Lett., 735, L13,
  \dodoi{10.1088/2041-8205/735/1/L13}

\bibitem[{Weidner {et~al.}(2010)Weidner, Kroupa, \& Bonnell}]{Weidner2010}
Weidner, C., Kroupa, P., \& Bonnell, I.~A. 2010, Mon. Not. R. Astron. Soc.,
  401, 275, \dodoi{10.1111/j.1365-2966.2009.15633.x}

\bibitem[{Weidner {et~al.}(2014)Weidner, Kroupa, \&
  Pflamm-Altenburg}]{Weidner2014}
Weidner, C., Kroupa, P., \& Pflamm-Altenburg, J. 2014, Mon. Not. R. Astron.
  Soc., 441, 3348, \dodoi{10.1093/mnras/stu640}

\bibitem[{Weisz {et~al.}(2012)Weisz, Johnson, Johnson, Skillman, Lee,
  Kennicutt, Calzetti, {Van Zee}, Bothwell, Dalcanton, Dale, \&
  Williams}]{Weisz2012}
Weisz, D.~R., Johnson, B.~D., Johnson, L.~C., {et~al.} 2012, Astrophys. J.,
  744, 44, \dodoi{10.1088/0004-637X/744/1/44}

\bibitem[{Weisz {et~al.}(2015)Weisz, Johnson, Foreman-Mackey, Dolphin, Beerman,
  Williams, Dalcanton, Rix, Hogg, Fouesneau, Johnson, Bell, Boyer, Gouliermis,
  Guhathakurta, Kalirai, Lewis, Seth, \& Skillman}]{Weisz2015}
Weisz, D.~R., Johnson, L.~C., Foreman-Mackey, D., {et~al.} 2015, Astrophys. J.,
  806, 198, \dodoi{10.1088/0004-637X/806/2/198}

\bibitem[{Wilkins {et~al.}(2008)Wilkins, Hopkins, Trentham, \&
  Tojeiro}]{Wilkins2008}
Wilkins, S.~M., Hopkins, A.~M., Trentham, N., \& Tojeiro, R. 2008, Mon. Not. R.
  Astron. Soc., 391, 363, \dodoi{10.1111/j.1365-2966.2008.13890.x}

\bibitem[{Yan {et~al.}(2017)Yan, Jerabkova, \& Kroupa}]{Yan2017}
Yan, Z., Jerabkova, T., \& Kroupa, P. 2017, Astron. Astrophys., 607, 1,
  \dodoi{10.1051/0004-6361/201730987}

\end{thebibliography}
\bibliographystyle{aasjournal}

\end{document}